\documentclass[reprint,nofootinbib, amsmath,amssymb, aps]{revtex4-2}
\usepackage{graphicx}
\usepackage{epstopdf}
\usepackage{dcolumn}
\usepackage{bm}
\usepackage{slashed}
\usepackage{amsmath}
\usepackage{amssymb}
\usepackage{xcolor}
\usepackage{mathrsfs}
\usepackage{subfigure}
\usepackage{hyperref}

\begin{document}

\title{Estimation of the cumulant $b_4$ and sextic self-coupling of the QCD axion at finite temperature and density}

\author{Xinguang Li}
\author{Zhao Zhang}
\email{zhaozhang@pku.org.cn}

\affiliation{School of Mathematics and Physics, North China Electric Power University, Beijing 102206, China}

\begin{abstract}

The sixth-order cumulant $b_4$ of the QCD topological charge distribution and the axion's sextic self-coupling at finite temperature
$T$ and baryon chemical potential $\mu$ are calculated within the two-flavor quark-meson (QM) model without and with the Polyakov-loop 
dynamics. The $b_4$ in the vacuum falls within the range predicted by lattice SU(3) pure-gauge simulations, and at large $T$ and/or $\mu$, 
it tends to a constant predicted by the dilute instanton gas approximation. Both $b_4$ and the axion's sextic self-coupling diverge at the 
QCD critical point and flip sign across the phase transition. The range of accuracy of the fourth- and sixth-order Taylor expansions of 
the free energy as a function of the $\theta$ parameter is investigated via comparison with the full potential at the mean-field level.                    

\end{abstract}

                              
\maketitle

\section{Introduction}

The true ground state of QCD is believed to be a superposition of an infinite number of 
degenerate vacuum states labeled by a topological winding number $n$, namely the $\theta$-vacuum \cite{Belavin:1975,Callan:1976,Jackiw:1976pf} 
\begin{equation}\label{Eq:thetq-vacuum}
|\theta \rangle=\sum_{n=-\infty}^{+\infty}  {e^{-i n \theta}} |n \rangle, 
\end{equation}
where $\theta$ is a real angle. The angular parameter $\theta$ enters the QCD Lagrangian through a total derivative 
term 
\begin{equation}\label{Eq:theta-term}
{\cal L}_{\theta}=\theta q(x), 
\end{equation}
where
\begin{equation}\label{Eq:tl-density}
q(x)= \frac{g^2}{32\pi^2} G^a_{\mu\nu}(x) \cdot\tilde{G}^{a\mu\nu}(x)
\end{equation}
is the topological charge density and $G^a_{\mu\nu}$ and $\tilde{G}^{a\mu\nu}$ denote the gluon field strength tensor 
and its dual respectively. Note that the $\theta$-term \eqref{Eq:theta-term} breaks the CP symmetry explicitly. The experimental bounds 
from the neutron electric dipole moment indicate that $\theta$ is extremely small: $|\theta| \lesssim 10^{-10}$\cite{1979Chiral,Baker:2006ts,Griffith:2009,Parker:2015yka,Graner:2016,Guo:2015tla}, 
which gives rise to the well-known strong CP problem.

An elegant solution to this problem is the Peccei–Quinn mechanism which promotes $\theta$ from a constant to a field by 
introducing an additional chiral $U(1)$ symmetry ($U(1)_{PQ}$) \cite{Peccei:1977hh,Peccei:1977ur}. Upon the spontaneous symmetry 
breaking of $U(1)_{PQ}$ at high-energy, this mechanism yields a pseudo-Goldstone boson, namely the QCD axion \cite{Weinberg:1977ma,Wilczek:1977pj}, 
whose anomalous coupling to gluons generates an effective potential that dynamically cancels the $\theta$-parameter in QCD \cite{kim1979weak,shifman1980can,dine1981simple,zhitnitsky1980possible,Kim:2008hd}. 
As a weakly interacting particle, the QCD axion is also a compelling candidate of the cold dark matter 
\cite{preskill1983cosmology,abbott1983cosmological,DINE1983137,KIM19871,Duffy:2009ig,Kawasaki:2013ae,Marsh:2015xka,DiLuzio:2020wdo,Bramante:2023djs}.
The corresponding axion search methods can been found in Refs. \cite{Graham:2015ouw,Irastorza:2018dyq,Sikivie:2020zpn,Semertzidis:2021rxs,Hook:2017psm}.  
  
The topological charge is defined as       
\begin{equation}\label{tl-charge}
Q= \int {dx^4} q(x)
\end{equation}
which is an integer characterized the distinct $\theta$-vacuum sector. The cumulants of the topological charge distribution 
are fundamental, non-perturbative probes of the QCD vacuum's topological structure and the properties of axion, which are
defined as
\begin{equation}
C_{2n}=(-1)^{n+1} \frac{d^{2n}}{d\theta^{2n}} F(\theta,T,\mu)|_{\theta=0}=\langle{Q^{2n}}\rangle_{c},
\end{equation}
where $F(\theta,T,\mu)$ is the free energy density. The most important cumulant is the topological susceptibility $\chi_t$,
which quantifies how strongly the QCD vacuum fluctuates between topologically distinct field configurations. 
This quantity has been studied extensively both at vacuum and finite temperature and density, which is directly related 
to the mass of the QCD axion through the relation $m_a^2=\chi_t/f_a^2$, where $f_a$ is the axion decay constant.
Cosmological and astrophysical observations place constraints on the QCD axion mass or decay constant \cite{turner1990windows,1990Astrophysical,DEramo:2022nvb,Notari:2022ffe,Raffelt:2006cw,Caputo:2024oqc,Sedrakian:2015krq,Choi:2020rgn}.  

The Taylor expansion of $F(\theta,T,\mu)$ can be rewrote as \cite{Vicari:2008jw}
\begin{equation}
F(\theta,T,\mu)-F(0,T,\mu)=\frac{1}{2} \chi_t(T,\mu) \theta^2 s(\theta,T,\mu) \label{eq:tailer1}
\end{equation}
and the dimensionless function $s(\theta,T,\mu)$ takes the form  
\begin{equation}
s(\theta,T,\mu)=1+b_2(T,\mu)\theta^2+b_4(T,\mu)\theta^4+\cdots, \label{eq:tailer2}
\end{equation}
where the coefficients $b_n$ are the normalized cumulants of the topological charge distribution.
The cumulants $b_2$ and $b_4$, which quantifies the non-Gaussian features of $ F(\theta,T,\mu)$, have also been investigated 
in the previous studies both at vacuum and medium. Note that the free energy $F(\theta,T,\mu)$ corresponds to the axion potential 
$V(a,T,\mu)$ when the angle $\theta$ is replaced by the axion field $a/f_a$. Similarly, the axion potential can be expanded as
\begin{equation}
V(a,T,\mu)=\frac{1}{2}m_a^2 a^2 +\frac{1}{4!}\lambda_{a,4}(T,\mu)a^4+\frac{1}{6!}\lambda_{a,6}(T,\mu) a^6+\cdots ,
\end{equation}
where $\lambda_{a,n}$ denotes the $n$th-order axion self-coupling. Clearly, the cumulants $b_2$ and $b_4$ are proportional to the 
quartic and sextic axion self-couplings, respectively.

Studying the nontrivial $\theta$-dependence of the ground-state energy density $F(\theta)$ provides valuable insight into both 
the topological structure of the QCD vacuum and the properties of the QCD axion. The lattice QCD simulations at zero baryon chemical 
potential provide the first principle results for the topological susceptibility \(\chi_t\) and the fourth-order topological cumulant \(b_2\) 
\cite{Bonati:2015vqz,Borsanyi:2015cka,Berkowitz:2015aua,Borsanyi:2016ksw,Petreczky:2016vrs,Bonati:2018blm,Athenodorou:2022aay,Kotov:2025ilm}. 
The topological susceptibility decreases monotonically across the chiral crossover temperature \(T_c\), while \(b_2\) stays negative and 
slowly approaches the dilute instanton gas limit at high temperatures \cite{1981QCD,Bonanno:2025wcv}. Technically, it is hard to evaluate the sixth-order cumulant \(b_4\)  
in lattice QCD and only a rough calculation of \(b_4\) is performed within the pure SU(N) lattice gauge theory \cite{Panagopoulos:2011rb,Bonati:2013tt,Bonati:2015sqt, Bonati:2016tvi,Bonanno:2018xtd,Bonanno:2020hht}.
So far, all lattice computations are restricted to zero baryon density due to the notorious sign problem. On the other hand, 
the chiral perturbation theory (ChPT) has been extensively used to study the topological susceptibility and higher topological cumulants 
\cite{GrillidiCortona:2015jxo,Guo:2015oxa,Gorghetto:2018ocs,Lu:2020rhp,Balkin:2020dsr,Springmann:2024mjp}, with its predictions frequently benchmarked against 
lattice QCD computations. Comparing to lattice simulations, the results obtained from ChPT are only reliable at low temperatures and 
densities.     

Recently, several studies based on the Nambu-Jona-Lasinio (NJL) models \cite{Nambu:1961A,Nambu:1961B,Klevansky:1992,Hatsuda:1994pi,BUBALLA:2005205} suggest that
the fourth cumulant $b_2$ or the axion quartic self-coupling $\lambda_{a,4}$ is enhanced significantly near the chiral critical point of QCD, even though it 
decreases markedly in the chiral symmetry-restored phase \cite{Lu:2018ukl,Das:2020pjg,Gong:2024cwc,Zhang:2023lij,Abhishek:2020pjg,Murgana:2024djt,Zhang:2025lan}. 
This is distinctly different from the behaviour of the topological susceptibility near the critical point, a quantity that always decreases with 
diminishing magnitude of the chiral condensate. First, this means that the no-Gaussian deviation of the topological charge distribution becomes 
significantly near the chiral critical point. Second, such a remarkable strengthening of $\lambda_{a,4}$ near the critical point may imply the 
potential for the formation or enhancement of an axion Bose-Einstein condensate (BEC) \cite{Sikivie:2009qn} inside the compact star. In addition, 
the thermal behavior of the sixth-order cumulant $b_4$ has been calculated via the two-flavor ChPT \cite{Lu:2026fhg}, though such results are 
only restricted to the low-temperature regime. 

A natural question arises as to whether the sixth-order cumulant or the sextic self-coupling of the axion exhibits divergent or rapidly 
rising behavior near the QCD critical point, which may have important implications for axion physics. On the other hand, it is also interesting 
to check the range of accuracy of the Taylor expansion of $F(\theta,T,\mu)$, once the lowest-order cumulants such as $C_2$, $C_4$, and $C_6$ have 
been calculated. This is a difficult task in the lattice calculation because of the sign problem for nonzero $\theta$ (also for nonzero $\mu$). 
However, it can be investigated in the effective theory or low-energy effective models of QCD, where the free energy at finite $\theta$ 
can be calculated directly. Studying the Taylor expansion approximation of the free energy as a function of $\theta$ using effective 
models helps to reliably assess the range of accuracy of lattice QCD results.

In this paper, we aim to estimate the sixth-order cumulant of the topological charge distribution at finite temperature and density,
particularly its behavior near the QCD critical point, by employing a low-energy effective model of QCD. Furthermore, we will study 
the range of accuracy of $F(\theta,T,\mu)$ as a function of the $\theta$ angle up to sixth-order in the Taylor expansion. The two-flavor quark 
meson (QM) model and its Polyakov-loop extended version (PQM), which have been extensively used in the investigation of QCD phase transition 
\cite{Schaefer:2007zz,Mizher:2009zz,Boomsma:2009eh,Skokov:2010sf,Zhang:2021zz,Gupta:2011ez,Rai:2022wth,Rai:2023tk}, will 
be adopted in our calculation. This model has the advantage of being renormalizable and is well suited for describing QCD phase 
transitions. We compare results obtained from quark-meson-type models with those from NJL-type models to investigate the model 
dependence of our conclusions. 

The rest of this paper is organized as follows. In section \ref{sect:1}, we present the general relations between the cumulants and the axion 
properties. In section \ref{sect:formalism}, we describe the (P)QM model at finite $\theta$ and show the method to calculate the topological 
observables within this formalism. In section \ref{sect:results}, we demonstrate the numerical results and provide relevant discussions. The 
conclusion and outlook are presented in section \ref{sect:conclusion}.

\section{ Cumulants $b_2$ and $b_4$, and their relations to axion properties } \label{sect:1}

In QCD, the normalized cumulants $b_2$ and $b_4$ of the topological charge distribution quantify the deviation of the topological 
charge distribution from the purely Gaussian behavior. They are central to understanding the non-perturbative structure of the QCD vacuum. In 
addition, the cumulants $b_2$  and $b_4$ are directly related to the quartic and sextic self-couplings of the QCD axion, respectively. 

According to Eqs. \eqref{eq:tailer1} and \eqref{eq:tailer2}, the topological susceptibility $\chi_t$ and the two dimensionless cumulants $b_2$ 
and $b_4$ are defined as

\begin{eqnarray}
\chi_t = \left. \frac{d^2 \mathcal{F}\left( \theta ,T,\mu \right)}{d\theta^2} \right|_{\theta=0},
\end{eqnarray}
 
\begin{eqnarray}
b_2 = \frac{1}{12\chi _t}\left. \frac{d^4 \mathcal{F}\left( \theta ,T,\mu \right)}{d\theta ^4} \right|_{\theta =0} = \frac{c_4}{12\chi _t},
\end{eqnarray}
and
\begin{eqnarray}
b_4=\frac{1}{360\chi _t}\left. \frac{d^6 \mathcal{F}\left( \theta ,T,\mu \right)}{d\theta ^6} \right|_{\theta =0}=\frac{c_6}{360\chi _t},
\end{eqnarray}
respectively. Note that the dilute instanton gas approximation predicts that 
\begin{equation}
b_{2n} = \frac{2(-1)^n}{(2n+2)!}
\end{equation}
which is valid for high enough temperatures \cite{Bonanno:2025wcv}.

The axion mass, the quartic axion self-coupling $\lambda_{a,4}$, and the sextic axion self-coupling $\lambda_{a,4}$, are defined as 
\begin{equation}
m_a^2=\left. \frac{d^2V\left( a,T,\mu\right)}{da^2} \right|_{a=0} = \frac{c_2}{f_a^2} = \frac{\chi_t}{f_a^2}, \label{eq:massaxion}
\end{equation}

\begin{equation}
\lambda_{a,4}=\left. \frac{d^4V\left( a,T,\mu\right)}{da^4} \right|_{a=0} = \frac{c_4}{f_a^4}, \label{eq:sc4}
\end{equation}
and  
\begin{equation}
\lambda_{a,6}=\left. \frac{d^6V\left(a ,T,\mu\right)}{d a^6} \right|_{a =0} = \frac{c_6}{f_a^6}, \label{eq:sc6}
\end{equation}
respectively.

The axion mass arises from nonperturbative QCD instanton effects after the spontaneous breaking of the Peccei-Quinn symmetry. Its magnitude 
is tightly linked to the axion decay constant. The axion mass controls the coherence scale of axion dark matter, dictates the equilibrium and 
stability of self-interacting axion bound structures including oscillons and axion stars, and shapes axion transport and emission processes 
within dense astrophysical environments. 
 
Axion self-couplings play crucial roles in environments where the axion field attains large amplitudes or where gravity amplifies 
their effects. First, the formation and evolution of axitons/oscillons are predominantly governed by the balance between the quartic attractive 
self-coupling and the sextic repulsive self-coupling \cite{Cyncynates:2021rtf}. Second, the properties of axion stars are strongly dependent on the 
axion self-couplings \cite{Braaten:2019knj,Braaten:2015eeu,Chavanis:2017loo,Carenza:2024ehj}, even though such self-interactions remain globally negligible 
for the main part of axion minicluster \cite{hogan1988axion,Kolb:1993zz}. The maximum mass of a dilute axion star is determined by the competing balance 
among the attractive quartic self-interaction, gravitational force, and gradient pressure (the sixth-order self-coupling is generally insignificant). 
In contrast, the formation and lifetime of the dense axion star rely on both the quartic and sextic axion self-interactions, which dominate at large 
field amplitude $a/f_a \sim \pi$ \cite{Chavanis:2017loo}.

\section{Two-flavor (P)QM meson model with nonzero $\theta$} \label{sect:formalism}

\subsection{The QM model with nonzero $\theta$}

The QM model, or the linear sigma model coupled to quark fields, incorporates both the quark and meson degrees of freedom. The two-flavor 
QM model with nonzero $\theta$ parameter has been considered in Ref. \cite{Boomsma:2009eh} and the Lagrangian takes the form
\begin{equation}
\mathcal{L}_{QM} = \mathcal{L}_q+\mathcal{L}_m
\label{eq:Lq}
\end{equation}
where 
\begin{equation}
\mathcal{L}_q = \bar{\psi} \left[ i \partial\!\!\!/ - g \left( \sigma + i \gamma_5 \eta + \boldsymbol{a}_0 \cdot \boldsymbol{\tau} + i \gamma_5 \boldsymbol{\pi} \cdot \boldsymbol{\tau} \right) \right] \psi
\label{eq:Lq}
\end{equation}
and 
\begin{equation}
\begin{split}
\mathcal{L}_{m} = & \ \ \frac{1}{2} \text{Tr} (\partial_\mu \phi^\dagger \partial^\mu \phi) - \frac{m^2}{2} \text{Tr} (\phi^\dagger \phi) \\
& - \frac{\lambda_1}{4} [\text{Tr} (\phi^\dagger \phi)]^2 - \frac{\lambda_2}{4} \text{Tr} [(\phi^\dagger \phi)^2] \\
& + \frac{\kappa}{2} [e^{i\theta} \det(\phi) + e^{-i\theta} \det(\phi^\dagger)] \\
& + \frac{1}{2} \text{Tr} \left[ \frac{H}{\sqrt{2}} (\phi + \phi^\dagger) \right].
\end{split} \label{eq:Lm} 
\end{equation}
In \eqref{eq:Lm},  the chiral field $\phi$ is defined as
\begin{equation}
\phi = \frac{1}{\sqrt{2}}(\sigma + i \eta) + \frac{1}{\sqrt{2}}(\boldsymbol{a}_0  + i \boldsymbol{\pi}) \cdot \boldsymbol{\tau},
\end{equation}
where $\boldsymbol{\tau}$ are Pauli matrix in flavor space. Note that the $\theta$ parameters enters the Lagrangian through the 
instanton induced determinant interaction \cite{1986How} with the coupling constant $\kappa/2$, which violates the $U(1)_A$ global symmetry 
explicitly.
      
One can rewrite the Lagrangian \eqref{eq:Lm} in the form
\begin{equation}
\begin{split}
\mathcal{L}_{m} = & \frac{1}{2}(\partial_\mu \sigma \partial^\mu \sigma+\partial_\mu \eta \partial^\mu \eta +\partial_\mu  \boldsymbol{\pi} \partial^\mu \boldsymbol{\pi} \\ 
&+ \partial_\mu \boldsymbol{a}_0 \partial^\mu \boldsymbol{a}_0)-U_m
\end{split}\label{eq:lm2}
\end{equation}
where the potential $U_m$ reads
\begin{equation}\label{eq:ptmeson}
\begin{split}
U_m(\sigma,\eta,\boldsymbol{\pi},\boldsymbol{a}_0)= & \ \frac{m^2}{2} (\sigma^2 + \boldsymbol{\pi}^2 + \eta^2 +\boldsymbol{a}_0 ^2) \\
& - \frac{\kappa}{2} \cos \theta \ (\sigma^2 + \boldsymbol{\pi}^2 - \eta^2 -\boldsymbol{a}_0 ^2) \\
& + \kappa \sin \theta \ (\sigma \eta - \boldsymbol{\pi} \cdot \boldsymbol{a}_0 ) - H \sigma \\
& + \frac{1}{4} (\lambda_1 + \frac{\lambda_2}{2}) (\sigma^2 + \eta^2 + \boldsymbol{\pi}^2 + \boldsymbol{a}_0 ^2)^2 \\
& + \frac{\lambda_2}{2} (\sigma \boldsymbol{a}_0 + \eta \boldsymbol{\pi} + \pi \times \boldsymbol{a}_0 )^2 \,.
\end{split}
\end{equation}
For nonzero $\theta$, the vacuum expectation values of $\sigma$ and $\eta$, namely $\bar{\sigma}$ and $\bar{\eta}$, may be nonvanishing. 
In this case, the corresponding vacuum part of the potential \eqref{eq:ptmeson} takes the form
\begin{equation}
\label{eq:vptmeson}
\begin{split}
U_m^{vac} = & \frac{\lambda}{4} (\bar{\sigma}^2 - v_\theta^2)^2 - H \bar{\sigma} + \frac{\lambda}{4} (\bar{\eta}^2 - u_\theta^2)^2 \\
&+ \kappa \sin \theta \, \bar{\sigma} \bar{\eta} + \frac{\lambda}{2} \bar{\sigma}^2 \bar{\eta}^2 - \frac{\lambda}{4} (v_\theta^4 + u_\theta^4)
\end{split}
\end{equation}  
where
\begin{equation}
v_\theta^2 \equiv \frac{\kappa \cos \theta -m^2}{\lambda} \quad , \quad u_\theta^2 \equiv v_\theta^2 - \frac{2 \kappa}{\lambda} \cos \theta,  
\end{equation}
and \(\lambda\equiv\lambda_1+\lambda_2/2\). 

The six model parameters $m^2$, $\kappa$, $H$, $\lambda_1$, $\lambda_2$, and the coupling $g$, are determined by the physical values of the 
meson masses $m_{\sigma}$, $m_{\eta}$, $m_{\boldsymbol{\pi}}$, $m_{\boldsymbol{a}_0}$, the decay constant $f_{\pi}$,  and the constitute quark 
mass $m_q$, in the vacuum. For zero $\theta$, only $\bar{\sigma}$ is nonvanishing and the $\sigma$ field can be written as $\sigma=\bar{\sigma}+s$.  
Then, the vacuum fluctuation part of $U_m$ becomes   
\begin{equation}\label{eq:mflac}
\begin{split}
U_{m}^{fluc} = & \frac{1}{2} \left[ m_\pi^2\boldsymbol{\pi}^2 + m_\sigma^2 s^2 + m_\eta^2 \eta^2 + m_{\boldsymbol{a}_0 }^2 \boldsymbol{a}_0 ^2 \right]  \\
& + \left( \lambda_1 + \frac{1}{2} \lambda_2 \right) \bar{\sigma} s \left( s^2 + \boldsymbol{\pi}^2 + \eta^2 \right)  \\
& + \left( \lambda_1 + \frac{3}{2} \lambda_2 \right) \bar{\sigma} s \boldsymbol{a}_0 ^2 + \lambda_2 \bar{\sigma} \eta \boldsymbol{\pi} \cdot \boldsymbol{a}_0   \\
& + \left( \frac{\lambda_1}{4}  + \frac{\lambda_2}{8}  \right) \left( s^2 + \boldsymbol{\pi}^2 + \eta^2 + \boldsymbol{a}_0 ^2 \right)^2  \\
& + \frac{\lambda_2}{2}  \left[ \left( s \boldsymbol{a}_0  + \eta \boldsymbol{\pi} \right)^2 + \left( \boldsymbol{\pi} \times \boldsymbol{a}_0  \right)^2 \right].
\end{split}
\end{equation}
From Eq. \eqref{eq:mflac}, the tree level curvature masses of the mesons $\sigma,\eta,\boldsymbol{\pi},$ and $\boldsymbol{a}_0$ are given by 
\begin{align}
\begin{split}
m_{\sigma}^2 &= m^2 - \kappa + \frac{3}{2}(2\lambda_1 + \lambda_2)\bar{\sigma}^2, \\
m_{\eta}^2 &= m^2 + \kappa + (\lambda_1 + \frac{1}{2}\lambda_2)\bar{\sigma}^2,\\
m_{\boldsymbol{\pi}}^2 &= m^2 - \kappa + \frac{1}{2}(2\lambda_1 + \lambda_2)\bar{\sigma}^2, \\
m_{\boldsymbol{a}_0}^2 &= m^2 + \kappa + (\lambda_1 + \frac{3}{2}\lambda_2)\bar{\sigma}^2.
\end{split}
\end{align}
The model parameters are obtained as 
\begin{eqnarray}
\lambda_1 &=& \frac{m_{\sigma}^2+m_{\eta}^2-m_{a_0}^2-m_{\pi}^2}{2\bar{\sigma}^2}  \label{eq:lamda1} \\
\lambda_2 &=& \frac{m_{a_0}^2-m_{\eta}^2}{\bar{\sigma}^2}  \label{eq:lamda2} \\
m^2 &=& m_{\pi}^2+\frac{m_{\eta}^2-m_{\sigma}^2}{2}  \label{eq:mm} \\ 
\kappa &=& \frac{m_{\eta}^2-m_{\pi}^2}{2}  \label{eq:kappa} \\ 
g&=&\frac{m_q}{\bar{\sigma}}  \label{eq:g}
\end{eqnarray}
where $m_q$ is the constituent quark mass. 
The stationarity condition for the effective potential \eqref{eq:vptmeson} at $\theta=0$ gives
\begin{equation}
H=m_{\pi}^2 \bar{\sigma} \label{eq:H}
\end{equation}
and $\bar{\sigma}=f_{\pi}$. 
We will take the values $m_{\boldsymbol{\pi}}=138$ MeV,$m_{\sigma}=500~(600)$ MeV,$m_{\boldsymbol{a}_0}=980$ MeV, $m_{\eta}=574$ MeV, 
and $f_{\pi}=93$ MeV in our numerical calculations. Following \cite{Boomsma:2009eh}, $g=3.3$ is chosen in this paper.   

At finite temperature $T$ and chemical potential $\mu$, the mean-field thermodynamic grand potential of the QM model without considering 
both the thermal and quantum fluctuations of meson fields is given by
\begin{equation}
\Omega_{\text{MF}} = U_m^{vac}(\theta; \bar{\sigma},\bar{\eta}) + \Omega_{q\bar{q}} (T,\mu;\bar{\sigma},\bar{\eta}),
\end{equation}
where the quark one-loop vacuum contribution reads  
\begin{equation}
 \Omega_{q\bar{q}} (T,\mu;\bar{\sigma},\bar{\eta}) = \Omega_{q\bar{q}}^{vac}+\Omega_{q\bar{q}}^{T,\mu} \label{eq:vloop}
\end{equation}
with 
\begin{eqnarray}
\Omega_{q\bar{q}}^{vac} &=& -2N_c \sum_{f} \int{\frac{d^3p}{\left( 2\pi \right) ^3}} E_p^f \nonumber\\
                        &=& \mathbf{-2N_c \sum_{f} \int{\frac{d^4p}{\left( 2\pi \right) ^4}} ln(p_0^2+{E_p^f}^2)+K}
\label{eq:qvacuum} 
\end{eqnarray}
and   
\begin{equation}
\begin{split}
\Omega_{q\bar{q}}^{T} &= -2N_cT \sum_{f} \int{\frac{d^3p}{\left( 2\pi \right) ^3} \left\{\ln \left[ 1 + e^{-\left( E_p^f -\mu\right) /T} \right] \right.} \\
&\quad \left. + \ln \left[ 1 + e^{-\left( E_p^f +\mu\right) /T} \right] \right\}. \label{eq:qthermal}
\end{split}
\end{equation}
In Eqs. \eqref{eq:qvacuum} and \eqref{eq:qthermal},  $E_{p}^f = \sqrt{p^2 + m_f^2}$, with $m_f = g \sqrt{\bar{\sigma}^2 + \bar{\eta}^2}$. The index $f$ 
denotes the flavor, and $K$ is an infinite constant independent of the fermion mass and therefore is dropped.

The quark one-loop vacuum contribution \eqref{eq:qvacuum} is divergent and thus the regularization is needed. 
Using the dimensional regularization near four dimension, $d=4-2\epsilon$, one gets the $\epsilon$ zeroth order potential as 
\begin{eqnarray}
\Omega_{q\bar{q}}^{vac} = \frac{N_c}{16\pi^2}\sum_{f} m_f^4 \left[ \frac{1}{\epsilon}+\frac{3}{2}-\gamma_E-\ln(\frac{m_f^2}{4\pi\Lambda^2})\right], \label{eq:dimr}
\end{eqnarray}
where $\gamma_E$ is the Euler constant and $\Lambda$ is the renormalization scale. Adding a $\overline{\text{MS}}$ counterterm  
\begin{eqnarray}
\delta \mathcal{L} = \frac{N_cN_f}{16\pi^2}g^4(\sigma^2+\eta^2)^2 \left[ \frac{1}{\epsilon}+\ln(4\pi)-\gamma_E\right] \label{eq:counterterm}
\end{eqnarray}
to the Lagrangian of the QM model, one gets the renormalized quark vacuum contribution  
\begin{equation}
\Omega_{q\bar{q}}^{\text{reg}} = \frac{N_c}{(4\pi)^2} \sum_f m_f^4 \left[ \frac{3}{2} + \ln \left( \frac{\Lambda^2}{m_f^2} \right) \right].
\end{equation}
The vacuum grand potential then becomes 
\begin{equation}
\Omega^{\text{vac}}_{\Lambda}(\theta; \bar{\sigma},\bar{\eta}) =\Omega_{q\bar{q}}^{\text{reg}}(\bar{\sigma},\bar{\eta}) + U_{m}^{\text{vac}}(\theta; \bar{\sigma},\bar{\eta}), \label{eq:UmUq1}
\end{equation}
which is superficially renormalization scale dependent. 

Upon inclusion of the quark vacuum fluctuation, the dependences of meson curvature masses on the model parameters are modified \cite{Gupta:2011ez,Rai:2022wth,Rai:2023tk}.
Correspondingly, the model parameters fixed by the vacuum physical quantities encode the counterterm contributions. 
Such a parameter re-determination at $\theta=0$ and $T=0$ was carried out in Ref. \cite{Rai:2022wth} via calculations 
of the vacuum meson curvature masses and the vacuum pion decay constant performed with Eq. \eqref{eq:UmUq1} at $\bar{\eta}=0$.
It is found that the parameters $\lambda_2$ and $m^2$ are modified as follows:  
\begin{equation}
\lambda_2 \ = \ \frac{m_{a_0}^2-m_{\eta}^2}{\bar{\sigma}^2} - \frac{N_c g^4}{\pi^2} \ln \left( \frac{\Lambda^2}{g^2 f_\pi^2} \right),  
\end{equation}
\begin{equation}
m^2\ =\ m_{\pi}^2+\frac{m_{\eta}^2-m_{\sigma}^2}{2} - \frac{N_c g^4 f_\pi^2}{2\pi^2},  
\end{equation}
while the parameters $\lambda_1$, $\kappa$, and $H$ still remain the same forms as in Eqs. \eqref{eq:lamda1}, \eqref{eq:kappa}, and \eqref{eq:H}.
Upon substituting the modified model parameters into the expression for $U_{m}^{\text{vac}}(\bar{\sigma},\bar{\eta})$, Eq. \eqref{eq:UmUq1} is recast as
\begin{eqnarray}\label{eq:UmUq2} 
\Omega_{\text{QM}}^{\text{vac}}(\theta; \bar{\sigma},\bar{\eta}) & = & \frac{1}{2}m^2(\bar{\sigma}^2+\bar{\eta}^2)-\frac{1}{2}\kappa \cos \theta (\bar{\sigma}^2-\bar{\eta}^2)-H \bar{\sigma} \nonumber\\
                        & + & \frac{1}{4} (\lambda_1+\frac{1}{2}\lambda_{2s}-\frac{N_c g^4}{2 \pi^2} \ln (\frac{\Lambda^2}{g^2f_{\pi}^2}))(\bar{\sigma}^2+\bar{\eta}^2)^2  \nonumber \\
                        & + & \kappa \sin \theta \bar{\sigma} \bar{\eta} + \frac{N_c}{8{\pi}^2} m_q^4 \left[ \frac{3}{2} + \ln \left( \frac{\Lambda^2}{m_q^2} \right) \right] \nonumber \\
                        & = & \frac{1}{2}m^2(\bar{\sigma}^2+\bar{\eta}^2)-\frac{1}{2}\kappa \cos \theta (\bar{\sigma}^2-\bar{\eta}^2)-H \bar{\sigma} \nonumber\\
                        & + & \frac{1}{4} (\lambda_1+\frac{1}{2}\lambda_{2s}-\frac{3N_c g^4}{4 \pi^2} )(\bar{\sigma}^2+\bar{\eta}^2)^2 + \kappa \sin \theta \bar{\sigma} \bar{\eta}  \nonumber \\
                        & + & \frac{N_c}{8{\pi}^2} g^4 (\bar{\sigma}^2+\bar{\eta}^2)^2 \ln \left( \frac{f_{\pi}^2}{\bar{\sigma}^2+\bar{\eta}^2} \right), \label{eq:vacnoscale}
\end{eqnarray}            
where $\lambda_{2s}$ refers to Eq. \eqref{eq:lamda2}. We see the complete cancellation of the scale parameter $\Lambda$ in Eq. \eqref{eq:vacnoscale}.
Then, the total thermodynamical potential in the mean filed level will be written as
\begin{equation}
\Omega_{\text{QM}}(T, \mu, \theta;\bar{\sigma},\bar{\eta}) = \Omega_{\text{QM}}^{\text{vac}}(\theta; \bar{\sigma},\bar{\eta})+\Omega_{q\bar{q}}^{T}(T,\mu; \bar{\sigma},\bar{\eta}).   
\end{equation}
The condensates $\bar{\sigma}$ and $\bar{\eta}$ at finite $\theta$ are than obtained by solving the gap equations
\begin{equation}
\frac{\partial \Omega_{\text{QM}} }{\partial \bar{\sigma}} = \frac{\partial \Omega_{\text{QM}} }{\partial \bar{\eta}} =0.  \label{eq:gapeqsQM}
\end{equation}

\subsection{The PQM model with nonzero $\theta$}

The Polyakov loop operator \cite{Polyakov:1978} in QCD is a Wilson loop oriented in the temporal direction
\begin{equation}
L(\vec{x}) = \mathscr{P} \exp \left[ i \int_0^\beta d\tau A_0(\vec{x}, \tau) \right],
\end{equation}
where $\mathscr{P}$ denotes path ordering, $A_0(\vec{x}, \tau)$ is the temporal component of the Euclidean gauge field $A_\mu$, and 
$\beta = 1/T$. The Polyakov loop field $\Phi$ is defined as the thermal expectation value of the color trace of the Polyakov loop operator:
\begin{equation}
\Phi(\vec{x}) = \frac{1}{N_c} \langle \text{Tr}_c L(\vec{x}) \rangle, \qquad \bar{\Phi}(\vec{x}) = \frac{1}{N_c} \langle \text{Tr}_c L^\dagger(\vec{x}) \rangle
\end{equation}
which serves as the order parameter of deconfinement transition in the pure gauge theory. The Polyakov loop potential $\mathcal{U}(\varPhi, \bar{\varPhi}, T)$
is a model based on the degrees of freedom of the Polyakov loop field and its conjugate, which can successfully describe the thermal phase 
transition of the pure gauge theory \cite{Pisarski:2000eq}.   

Since the dynamical quarks explicitly breaks the $Z(3)$ center symmetry, the Polyakov loop is usually used as an approximate 
indicator for the confinement-deconfinement transition in QCD. The PQM model is a synthesis of the Polyakov loop model and the QM model by 
coupling the quark field to a constant background $SU(3)$ gauge field $A_\mu$. The quark confinement is partially mimicked in a statistical 
sense in this or similar type model, where the quark distribution function is modified by the small $\varPhi$ at low temperature.          

It should be noted that the primary focus of this work is not the deconfinement phase transition itself, but rather an assessment of whether 
the various topological observables calculated in the Polyakov loop enhanced chiral model demonstrate improved agreement with lattice QCD results.

The Lagrangian of the PQM at finite $\theta$  reads

\begin{eqnarray}
    \mathcal{L}_{\text{PQM}}
    &=& \bar{\psi} \left[ i\gamma_\mu D^\mu - g \left( \sigma + i \gamma_5 \eta + \boldsymbol{a}_0 \cdot \boldsymbol{\tau} + i \gamma_5 \boldsymbol{\pi} \cdot \boldsymbol{\tau} \right) \right] \psi \nonumber  \\
   &+& \mathcal{L}_m - \mathcal{U}\left( \varPhi ,\bar{\varPhi} ,T \right). \label{eq:lagforPQM}
\end{eqnarray}
where $D^\mu = \partial^\mu - i A^\mu$ with $A^\mu = \delta_{\mu 0} A^0$ the background SU(3) gauge field, $\mathcal{L}_m$ is the same as in Eq. \eqref{eq:Lm}, and $\mathcal{U}\left( \varPhi ,\bar{\varPhi} ,T \right)$ 
is the Polyakov loop potential. 

In this paper, we employ a simple polynomial model of $\mathcal{U}\left( \varPhi ,\bar{\varPhi} ,T \right)$ \cite{Pisarski:2000eq, Ratti:2005jh}, which takes the form  
\begin{equation}
\frac{\mathcal{U}\left( \varPhi ,\bar{\varPhi} ,T \right)}{T^4} = -\frac{b_2\left( T \right)}{2}\bar{\varPhi} \varPhi -\frac{b_3}{6}\left( \varPhi ^3+\bar{\varPhi} ^3 \right) +\frac{b_4}{4}\left( \varPhi \bar{\varPhi} \right) ^2
\label{eq:plpt}
\end{equation}
with
\begin{equation}
b_2\left( T \right) =a_0+a_1\left( \frac{T_0}{T} \right) +a_2\left( \frac{T_0}{T} \right) ^2+a_3\left( \frac{T_0}{T} \right)^3.
\end{equation}
The coefficients $a_i$, $b_i$, and the critical temperature $T_0$ are determined using pure-gauge lattice data \cite{Ratti:2005jh}. The values of $a_i$ and $b_i$ are provided in Table 1. 
In the pure gauge sector, the critical temperature $T_0$ for the confinement-deconfinement phase transition is set to 270 MeV. Correspondingly, in the present case 
with two dynamical flavors, we adopt a value of $T_0 = 208$ MeV \cite{Ratti:2005jh}.
\begin{table}[htbp]
\centering
\begin{tabular}{cccccc} 
\hline\hline
$a_0$~~~~ & $a_1$~~~~  &$ a_2 $~~~~ & $a_3$~~~~  & $b_3$~~~~ & $b_4$~~~~  \\ 
6.75~~~~ & -1.95~~~~ & 2.625~~~~ & -7.44~~~~ & 0.75~~~~ & 7.5~~~~   \\
\hline\hline
\end{tabular}
\caption{Parameters for the Polyakov loop potential.}
\end{table}

By adopting a gauge in which $A_0$ is constant, the Polyakov loop $L$ simplifies as a diagonal form in color space. The mean field
thermodynamical potential of the PQM is then given by 

\begin{eqnarray}
\Omega_{\text{PQM}} &= \Omega_{\text{QM}}^{\text{vac}}(T, \mu, \theta;\bar{\sigma},\bar{\eta})+\Omega_{q\bar{q}}^{T}(T,\mu; \bar{\sigma},\bar{\eta},\varPhi,\bar{\varPhi}) \nonumber \\
&+\mathcal{U}\left( \varPhi ,\bar{\varPhi} ,T \right), \label{eq:PQMmftp}
\end{eqnarray}
where 
\begin{equation}
\Omega_{q\bar{q}}^{T}=-2 T\sum_{f}\int\frac{d\boldsymbol{p}}{(2\pi)^3}\left(\ln g_f^{+}+\ln g_f^{-}\right), \label{19}
\end{equation}
with
\begin{align}
g_f^{+} &=  \left[ 1 + 3\varPhi e^{ -\beta \omega _{-}^{f} } + 3\bar{\varPhi} e^{ -2\beta \omega _{-}^{f} } + e^{ -3\beta \omega _{-}^{f} } \right],  \\
g_f^{-} &=  \left[ 1 + 3\bar{\varPhi}e^{ -\beta \omega _{+}^{f} } + 3\varPhi e^{ -2\beta \omega _{+}^{f} } + e^{ -3\beta \omega _{+}^{f} } \right]. 
\end{align}
In the above equations, $\omega _{\pm}^{f}=E_{p}^{f}\pm \mu$ refer to the single-particle excitation energies for antiquarks and quarks, respectively.
Note that the renormalization of the quark one-loop vacuum  contribution in the PQM is identical to that in the QM model. Consequently, the model 
parameters $\lambda_1$, $\lambda_2$, $m^2$, $H$, $g$ in the PQM model are also the same as those in the QM model.  

The condensates $\bar{\sigma}$, $\bar{\eta}$, $\varPhi$, and $\bar{\varPhi}$ at given $T$, $\mu$, and $\theta$ are than obtained by 
solving the following coupled equations
\begin{equation}
\frac{\partial \Omega_{\text{PQM}} }{\partial \bar{\sigma}} = \frac{\partial \Omega_{\text{PQM}} }{\partial \bar{\eta}} 
= \frac{\partial \Omega_{\text{PQM}} }{\partial \varPhi} = \frac{\partial \Omega_{\text{PQM}} }{\partial \bar{\varPhi}} = 0.  \label{eq:gapeqsPQM}
\end{equation}

\graphicspath{{figs/}}

\begin{figure}[!t]
\centering
\includegraphics[width=0.45\textwidth]{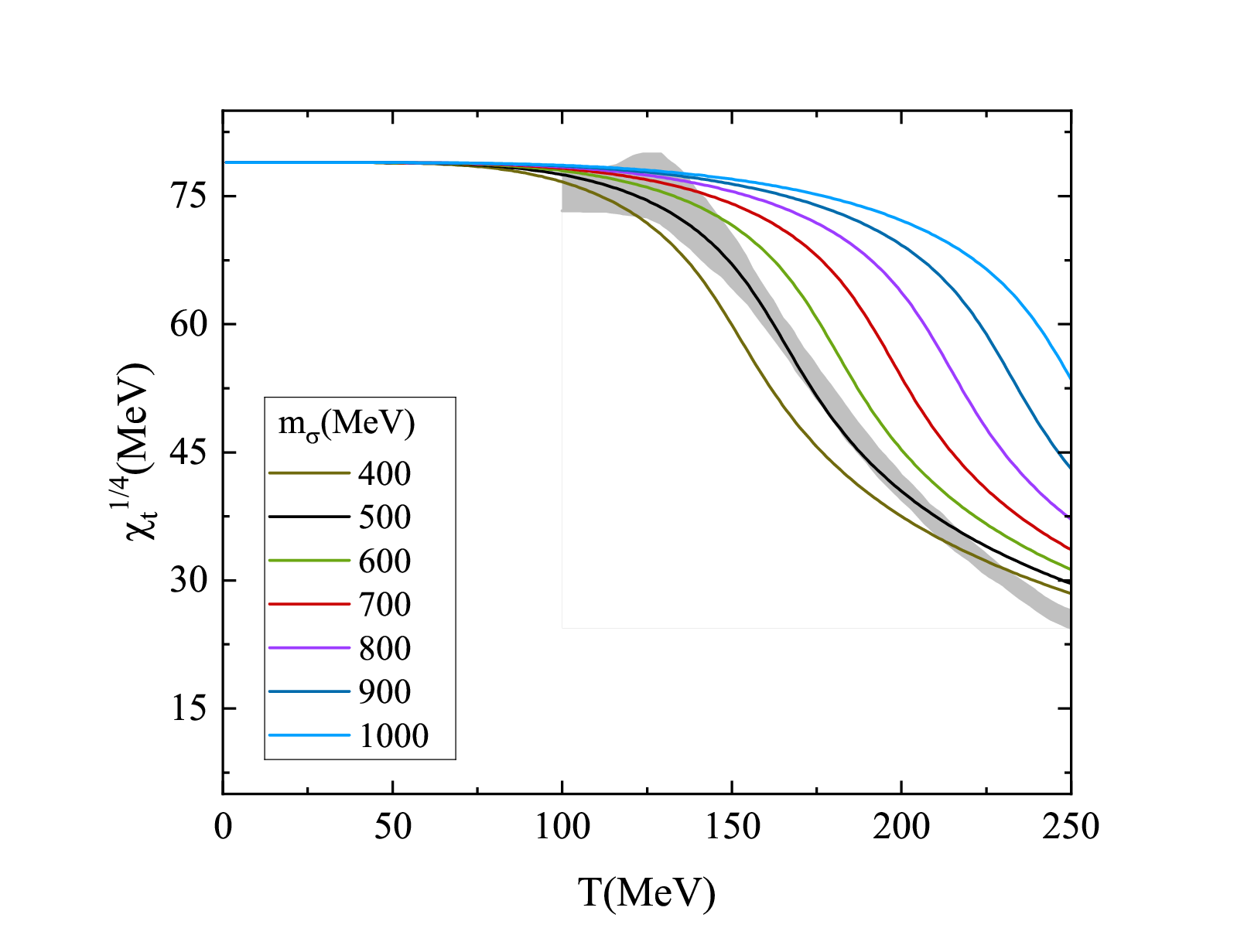}
\caption{The QCD topological susceptibility as a function of temperature for different $\sigma$ meson masses in the QM model. 
The shaded region in the figure represents the lattice QCD results \cite{Borsanyi:2016ksw}.}
\label{fig:topsus}
\end{figure}

\subsection{Evaluation of topological cumulants in (P)QM }

The QCD topological cumulants $c_2$ ($\chi_t$), $c_4$, and $c_6$ at finite $\theta$ is defined as 
\begin{eqnarray}
c_n = \frac{d^n \Omega_{\text{QCD}}}{d \theta^n}\Big|_{\theta=0}. 
\end{eqnarray}
In the (P)QM formalism, the cumulants can be evaluated via
\begin{eqnarray}
c_n = \frac{d^n V}{d \theta^n}\Big|_{\theta=0}, 
\end{eqnarray}
where $V$ is an effective potential of QCD axion \footnote{It corresponds to the potential of QCD axion if $\theta$ is replaced by the axion field $a/f_a$.} 
defined as 
\begin{equation}
V(T,\mu,\theta)=\Omega_{\text{(P)QM}}(x_i(T,\mu,\theta),T,\mu,\theta), \label{eq:AP}
\end{equation} 
and $x_i(T,\mu,\theta)$ refer to the physical values of $\bar{\sigma}$ and $\bar{\eta}$ ( $\bar{\sigma}$, $\bar{\eta}$, $\varPhi$ and $\bar{\varPhi}$), which are solutions of the 
gap equations \eqref{eq:gapeqsQM} (\eqref{eq:gapeqsPQM}) at given $T$, $\mu$, and $\theta$. Since the physical condensates are all implicitly dependent 
on $\theta$, the total differential of $V(\theta)$ with respect to $\theta$ satisfies the following relation
\begin{equation}
 \frac{\mathrm{d}{V}}{\mathrm{d} \theta}=\frac{\partial {V}}{\partial \theta}+\frac{\partial {V}}{\partial x_i }\frac{\partial x_i }{\partial \theta}. \label{eq:diffVa}
\end{equation}
Therefore to calculate $c_n$, we need to compute the 1-nth partial derivatives of each of the physical condensates 
with respect to $\theta$,  namely $\frac{\partial^{(m)}x_i}{\partial \theta^{(m)}}|_{\theta=0}$, where $m=1, \cdots, n$. 
This can be fulfilled by taking the successive derivatives of the gap equations \eqref{eq:gapeqsQM} or \eqref{eq:gapeqsPQM} with respect to $\theta$. 
First, the system of linear equations with $\frac{\partial x_i}{\partial \theta }$ as variables can be obtained by differentiating the gap equations \eqref{eq:gapeqsQM} 
or \eqref{eq:gapeqsPQM} with respect of $\theta$ and thus $\frac{\partial x_i}{\partial \theta}|_{\theta=0}$ can be calculated using the numerical solutions of \eqref{eq:gapeqsQM} 
or \eqref{eq:gapeqsPQM} at $\theta=0$. Second, the linear equations with $\frac{\partial^{(2)}x_i}{\partial \theta^{(2)}}$ as variables can be obtained by taking 
the second-order derivative of \eqref{eq:gapeqsQM} or \eqref{eq:gapeqsPQM} with respect to $\theta$ and then $\frac{\partial^{(2)}x_i}{\partial \theta^{(2)}}|_{\theta=0}$ 
can be computed (since $x_i|_{\theta=0}$ and $\frac{\partial x_i}{\partial \theta}|_{\theta=0}$ are all known). The higher order derivatives can be calculated in a similar way.    
Such a method has been used to evaluate the mass and self-coupling of QCD axion within the NJL framework \cite{Abhishek:2020pjg,Zhang:2025lan}. 

\begin{figure}[!t]
\centering
\begin{minipage}[h]{0.45\textwidth}
\centering
\includegraphics[width=1.0\textwidth]{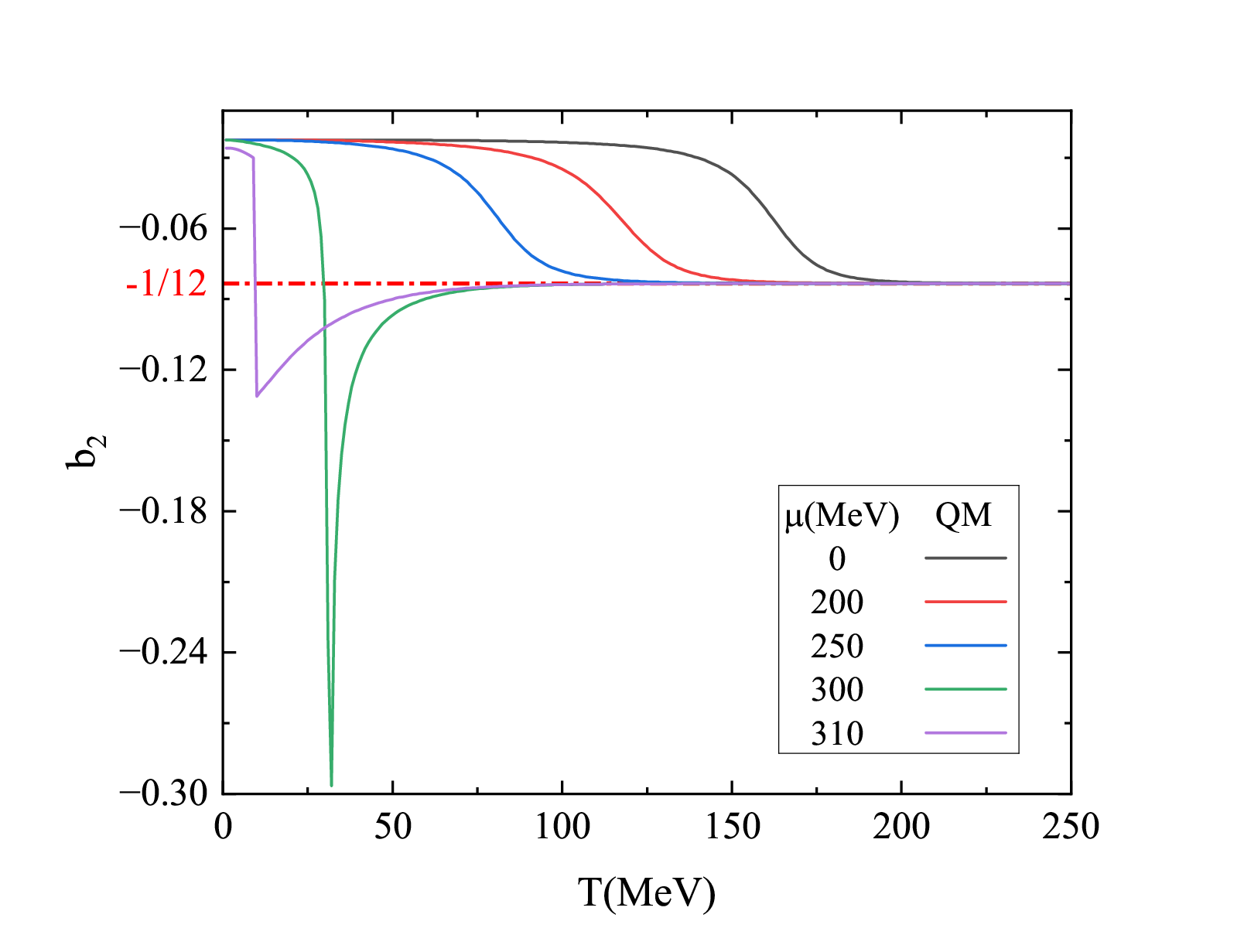}
\end{minipage}
\begin{minipage}[h]{0.45\textwidth}
\centering
\includegraphics[width=1.0\textwidth]{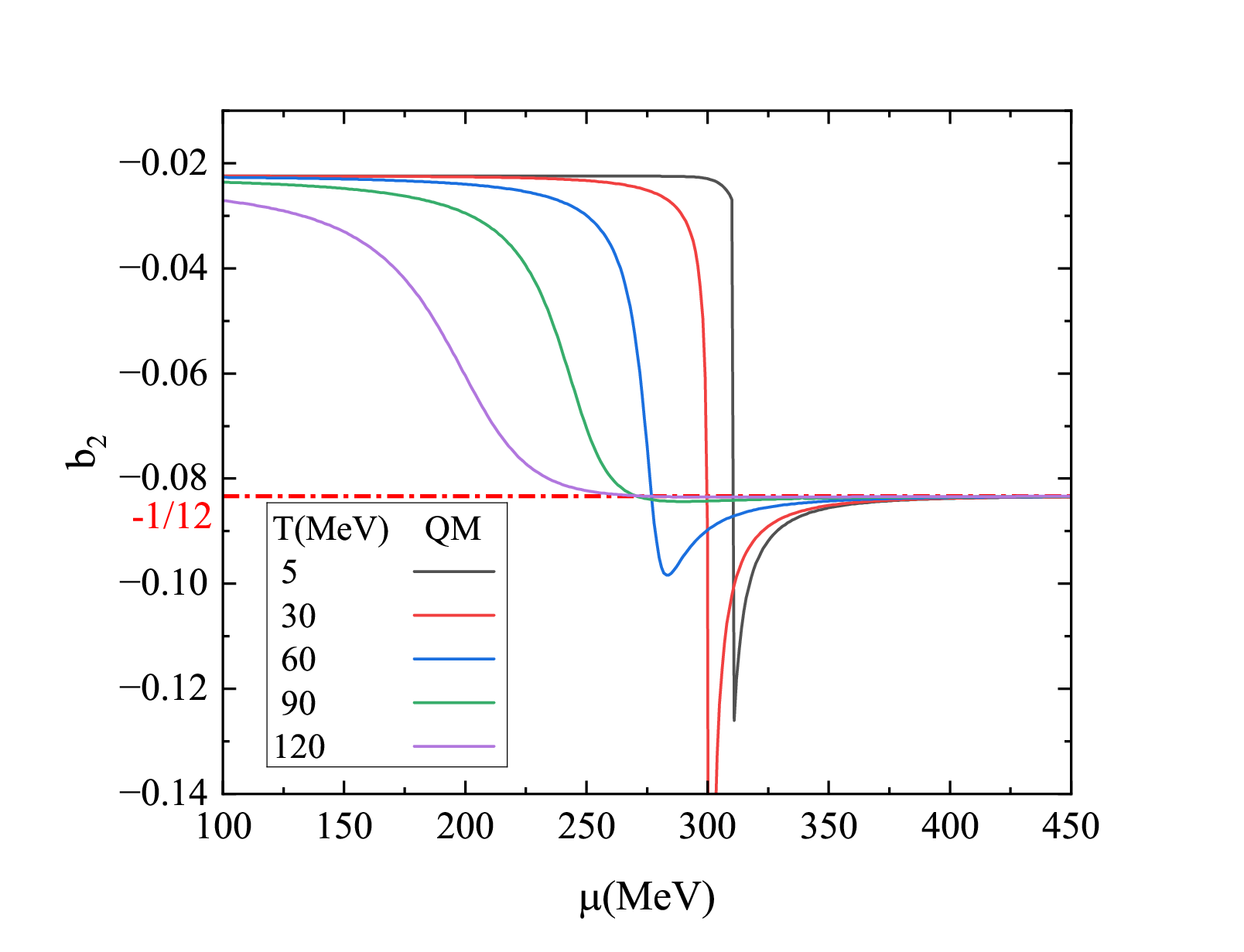}
\end{minipage}
\caption{The normalized fourth-order cumulant $b_2$ of the QCD topological charge distribution as a function of temperature at different quark chemical potentials (upper panel) 
and as a function of quark chemical potential at different temperatures (lower panel), obtained within the QM model. The red dash-dotted line represents the cumulant $b_2$ 
obtained from the dilute instanton gas approximation.  }
\label{fig:b2QM}
\end{figure}

\begin{figure}[!t]
\centering
\begin{minipage}[h]{0.45\textwidth}
\centering
\includegraphics[width=1.0\textwidth]{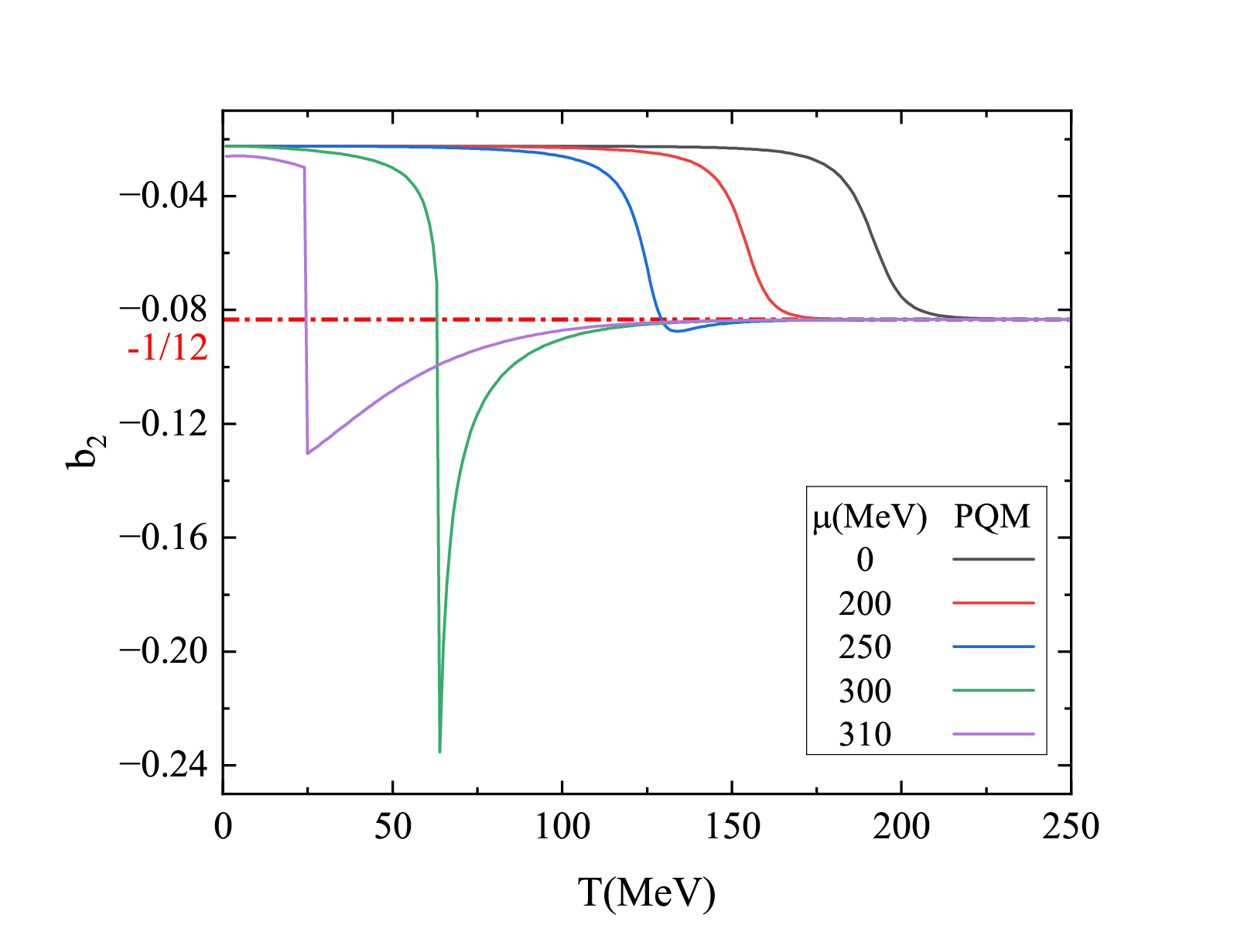}
\end{minipage}
\begin{minipage}[h]{0.45\textwidth}
\centering
\includegraphics[width=1.0\textwidth]{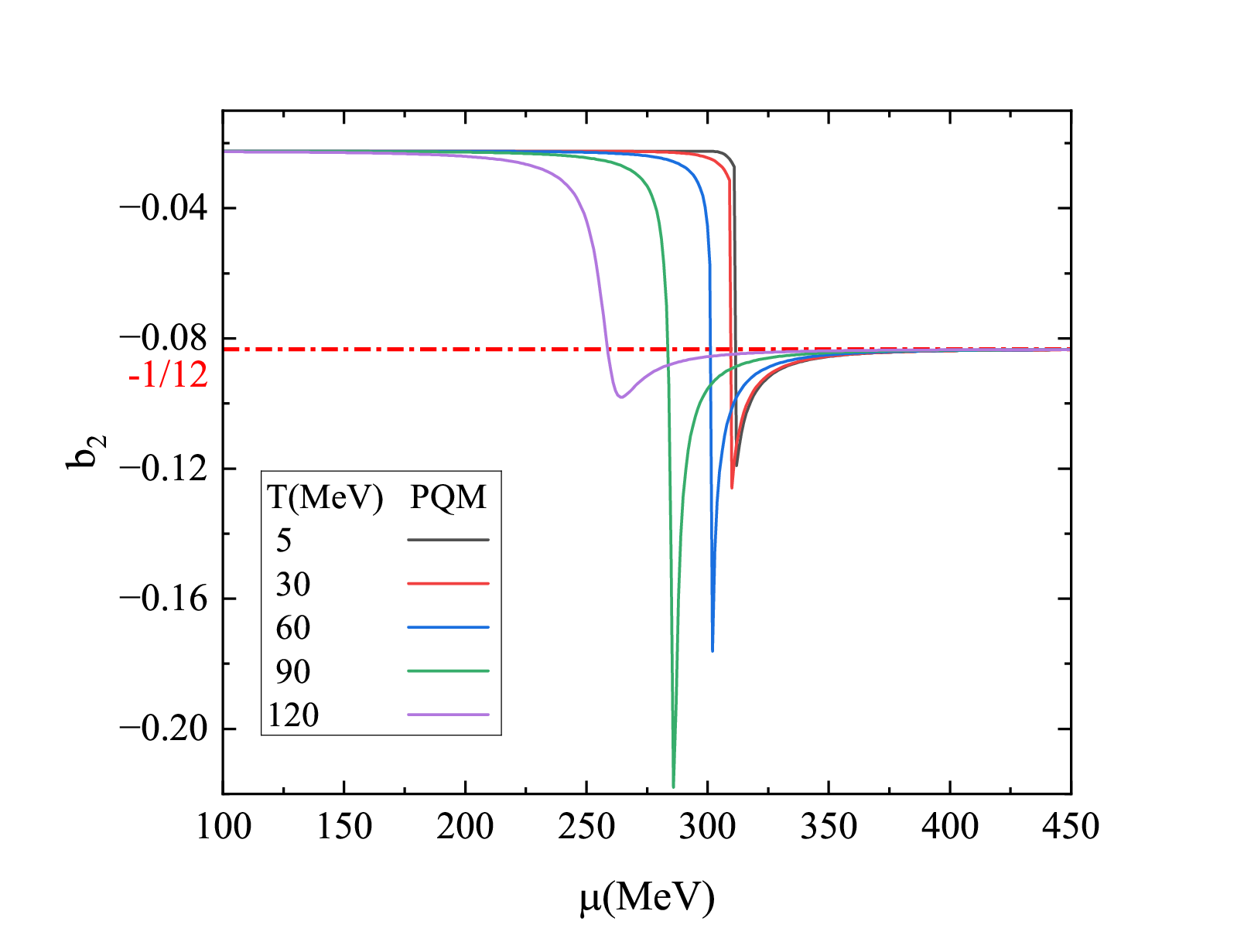}
\end{minipage}
\caption{Same as Fig. \ref{fig:b2QM}, but obtained in the PQM model.}
\label{fig:b2PQM}
\end{figure}

\section{\label{sect:results} NUMERICAL RESULTS AND DISCUSSIONS} 

\subsection{The topological observables and axion self-couplings at finite $T$ and $\mu$ }

\begin{figure}[!t]
\centering
\begin{minipage}[h]{0.45\textwidth}
\centering
\includegraphics[width=1.0\textwidth]{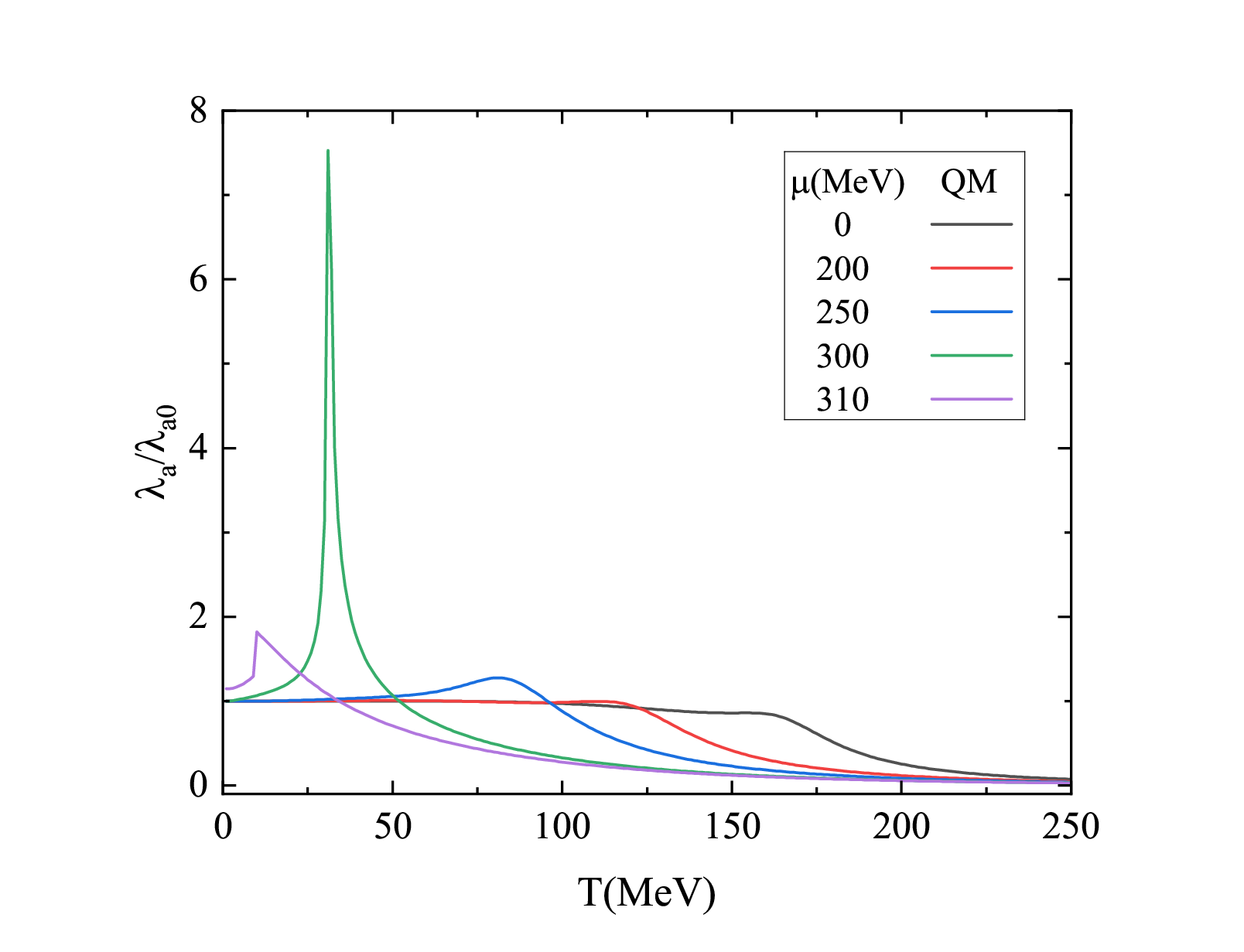}
\end{minipage}
\begin{minipage}[h]{0.45\textwidth}
\centering
\includegraphics[width=1.0\textwidth]{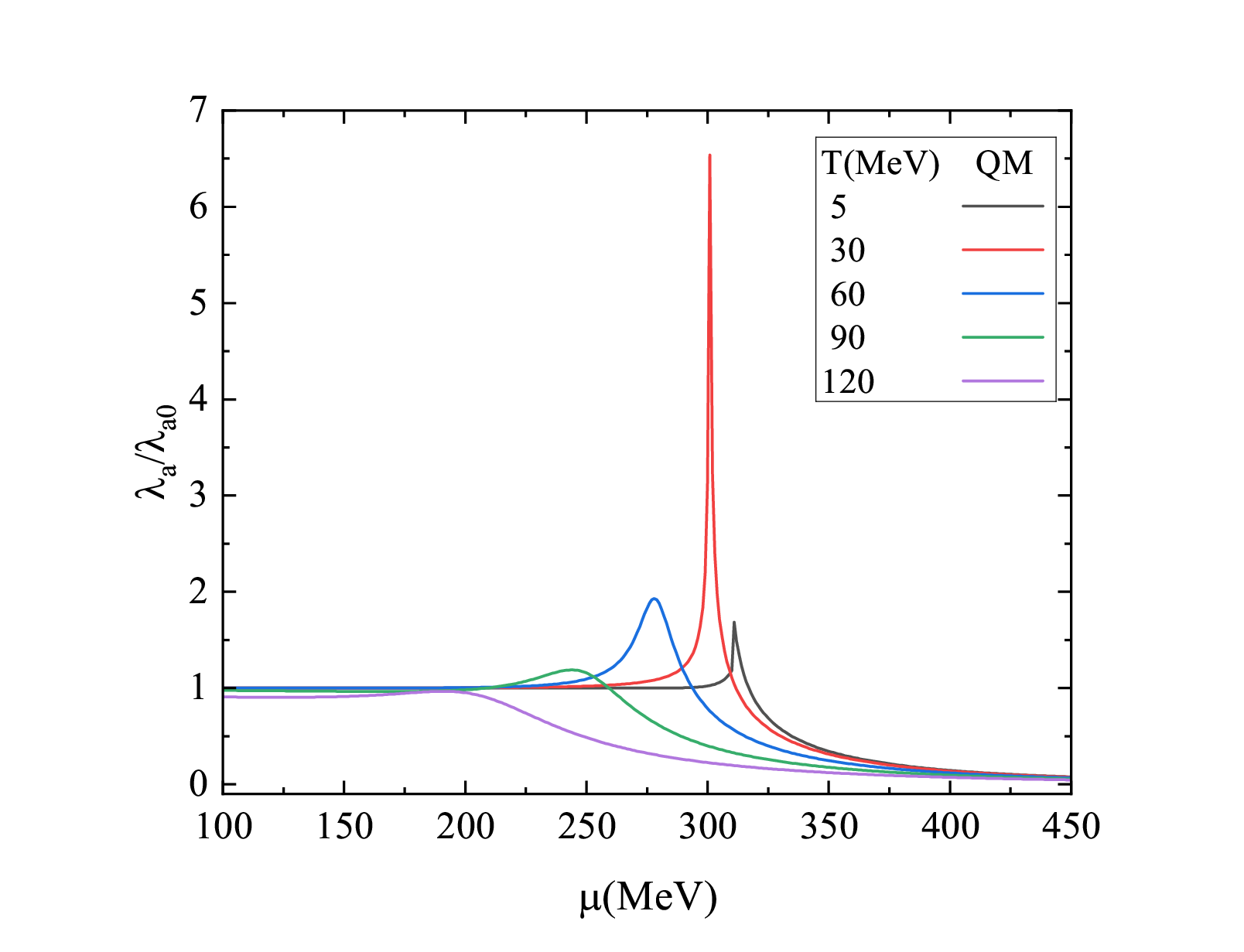}
\end{minipage}
\caption{The axion quartic self-coupling constant, scaled by its vacuum value, as a function of temperature at different quark chemical potentials (upper panel) and as a function of quark chemical potential at different temperatures (lower 
panel). All lines are obtained within the QM model.}
\label{fig:cp4QM}
\end{figure}

\begin{figure}[!t]
\centering
\begin{minipage}[h]{0.45\textwidth}
\centering
\includegraphics[width=1.0\textwidth]{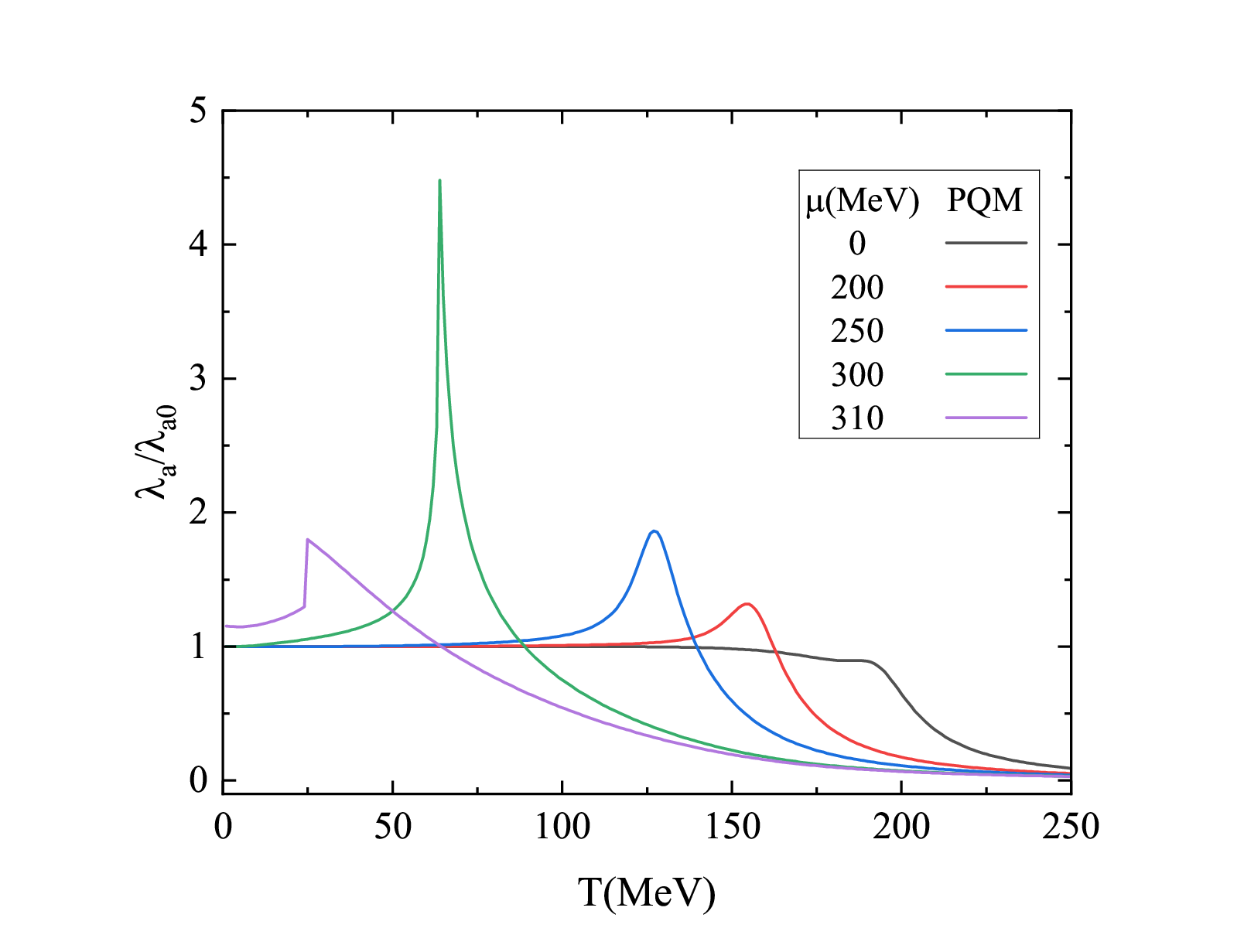}
\end{minipage}
\begin{minipage}[h]{0.45\textwidth}
\centering
\includegraphics[width=1.0\textwidth]{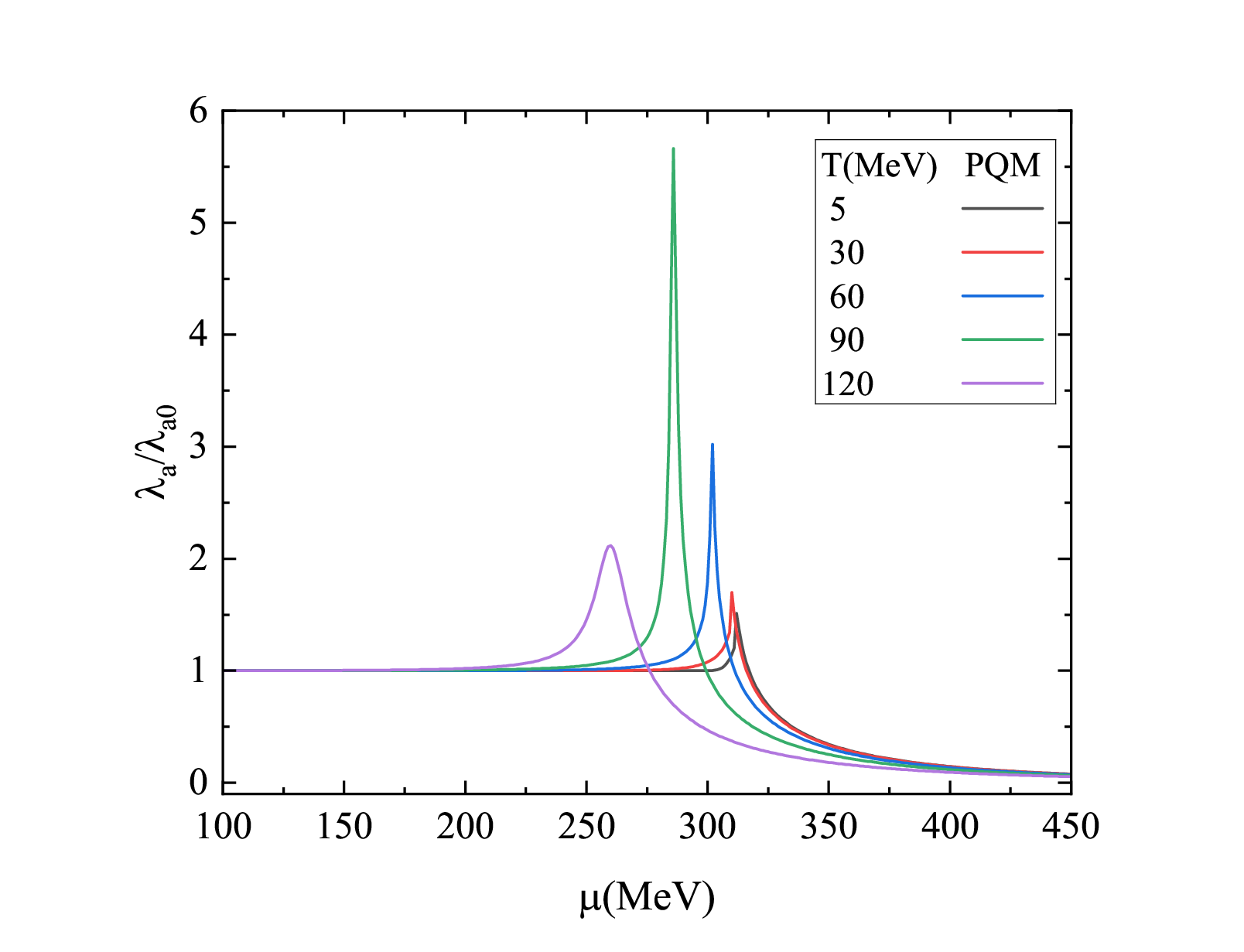}
\end{minipage}
\caption{Same as Fig. \ref{fig:cp4QM}, but obtained within the PQM model.}
\label{fig:cp4PQM}
\end{figure}

\begin{figure}[!t]
\centering
\begin{minipage}[h]{0.45\textwidth}
\centering
\includegraphics[width=1.0\textwidth]{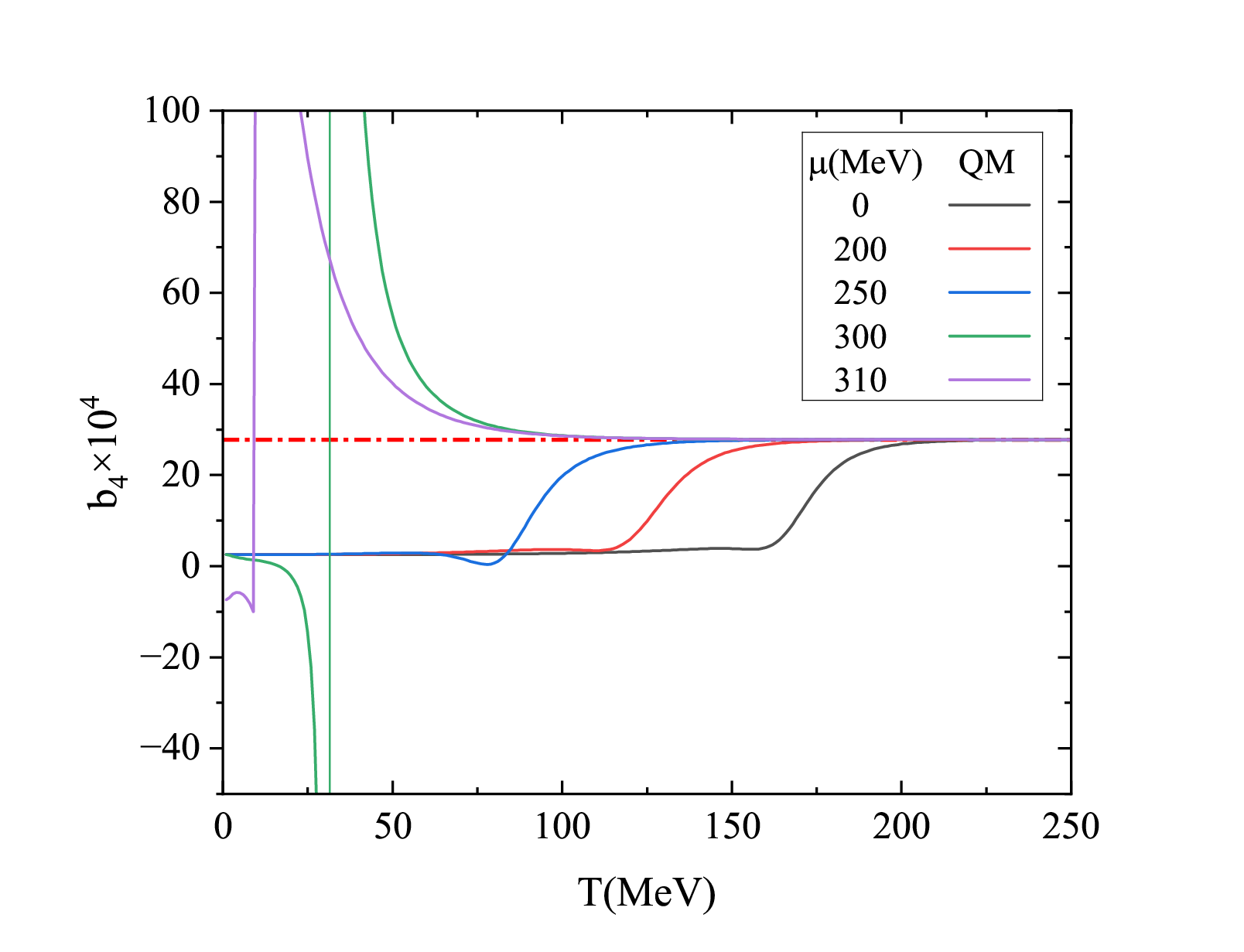}
\end{minipage}
\begin{minipage}[h]{0.45\textwidth}
\centering
\includegraphics[width=1.0\textwidth]{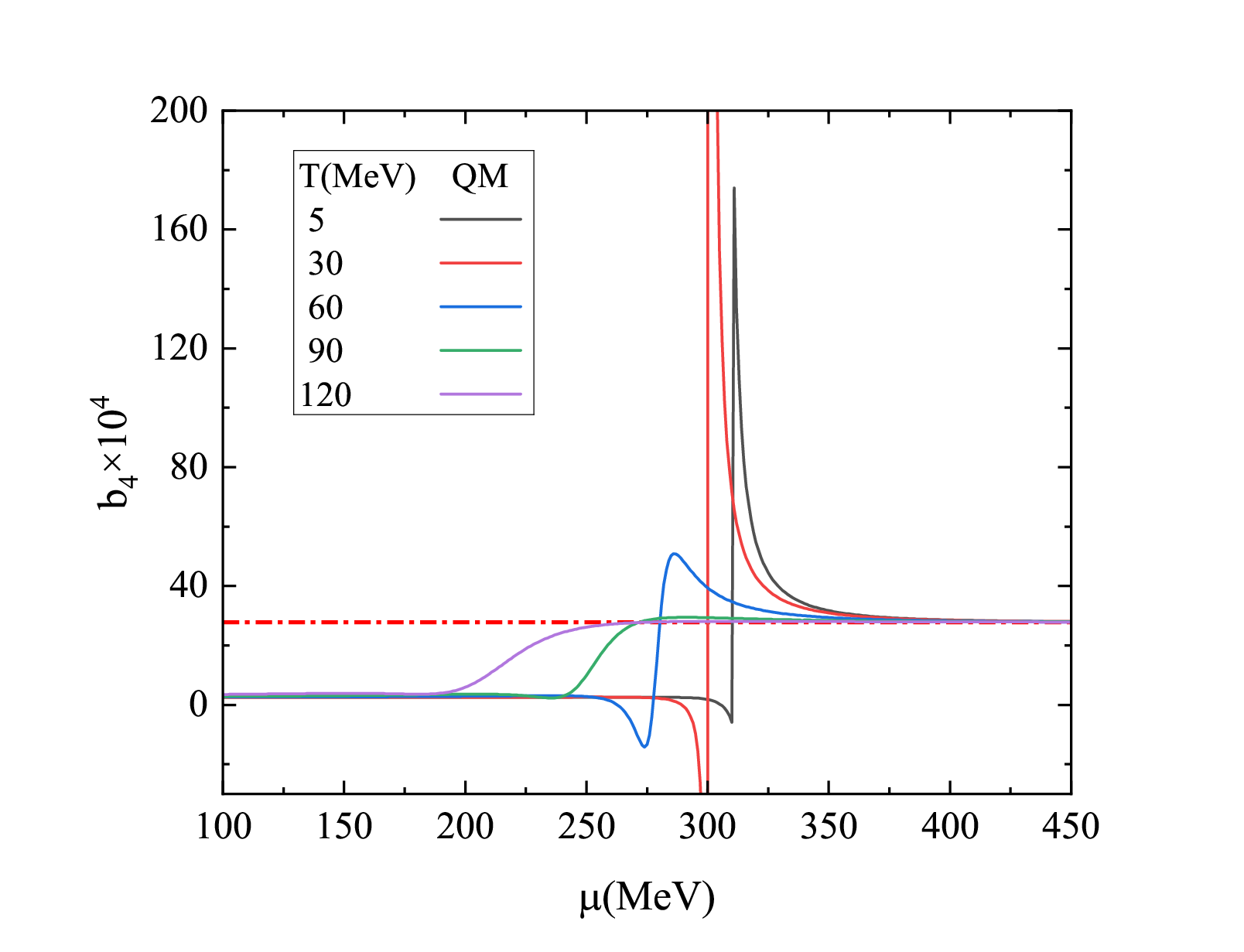}
\end{minipage}
\caption{The normalized sixth-order cumulant $b_4$ of the QCD topological charge distribution as the function of temperature at different quark chemical potentials (upper panel) 
and as the function of quark chemical potential at different temperatures (lower panel), obtained in the QM model. The red dash-dotted line represents ten thousand times the 
cumulant $b_4$ obtained from the dilute instanton gas approximation. }
\label{fig:b4QM}
\end{figure}

\begin{figure}[!t]
\centering
\begin{minipage}[h]{0.45\textwidth}
\centering
\includegraphics[width=1.0\textwidth]{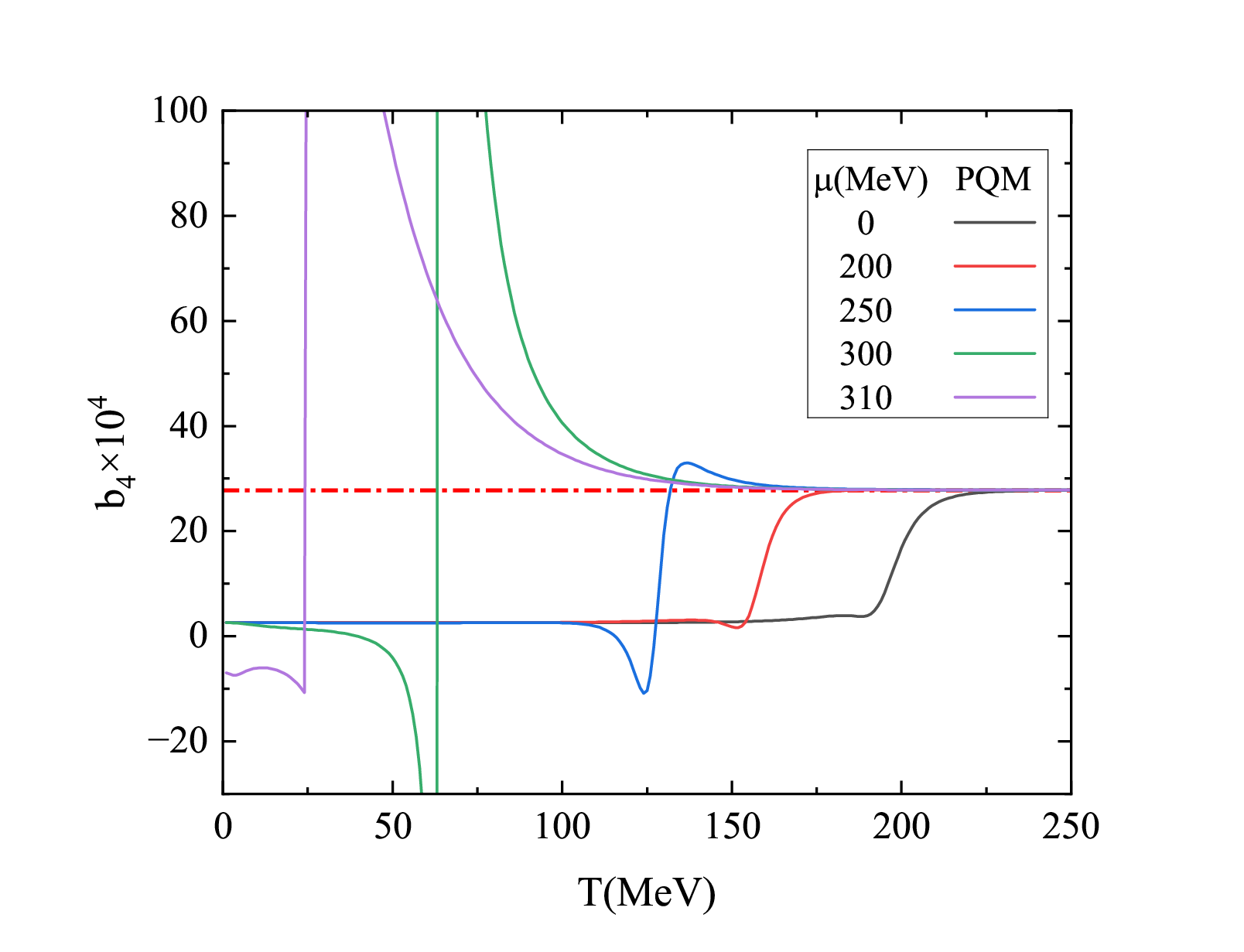}
\end{minipage}
\begin{minipage}[h]{0.45\textwidth}
\centering
\includegraphics[width=1.0\textwidth]{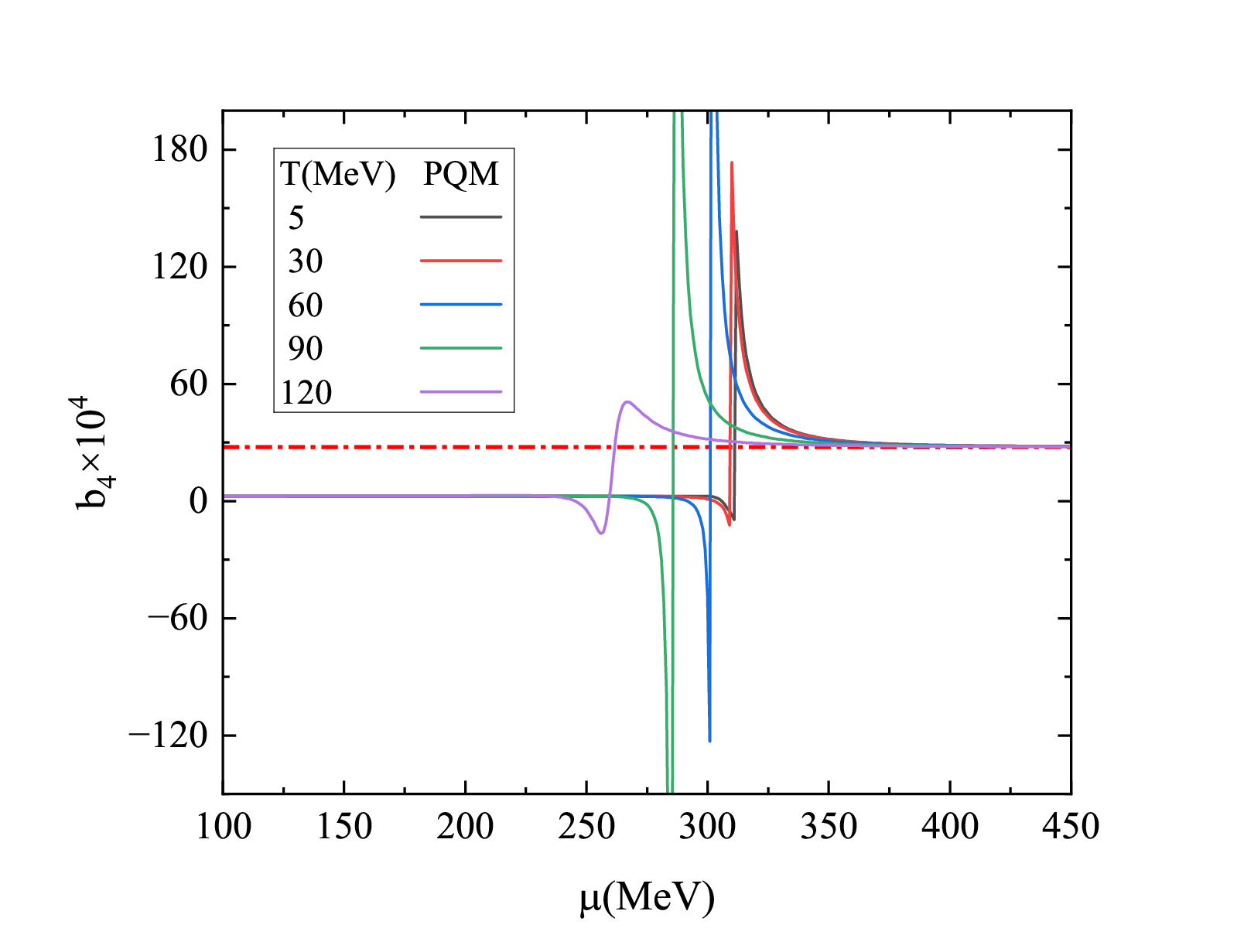}
\end{minipage}
\caption{Same as Fig. \ref{fig:b4QM}, but obtained within the PQM model.}
\label{fig:b4PQM}
\end{figure}

Figure \eqref{fig:topsus} shows the topological susceptibility as the function of the temperature for different $m_\sigma$ within 
the QM formalism. We see that for $m_\sigma$ in the range of 400–1000 MeV, the topological susceptibility at zero temperature takes 
a uniform value of $\chi_t^{1/4}=78.9~\text{MeV}$, which is very close to the ChPT result $\chi_t^{1/4}=77.8(4) \text{MeV}$ and
the lattice result $\chi_t^{1/4}=78.1(2) \text{MeV}$ for the two flavor case with the degenerate up-down quark masses \cite{GrillidiCortona:2015jxo,Borsanyi:2016ksw}. 
Fig \eqref{fig:topsus} indicates that the thermal behavior of $\chi_t$ is quite sensitive to $m_{\sigma}$ and the curve for $m_{\sigma}=500 \text{MeV}$ 
agrees well with the lattice result \cite{Borsanyi:2016ksw}.

The $T$-dependence of the normalized cumulant $b_2$ obtained in QM for different chemical potentials $\mu=0, 100, 200, 250, 300~\text{MeV}$ is 
shown in the upper panel of Fig. \ref{fig:b2QM}. This model predicts $b_2=-0.02246$ at zero $T$ and $\mu$, which agrees perfectly well with the 
ChPT result $b_2=-0.022(1)$ \cite{GrillidiCortona:2015jxo} for the isospin symmetric case.  Unlike the unchanged $\chi_t$, $b_2$ at vacuum 
increases with the value of $m_{\sigma}$ in QM, as indicated in Table \ref{222}. We see that at fixed lower quark chemical potentials, $b_2$ first decreases 
slightly with $T$, and then drops rapidly near the crossover temperature for the chiral restoration, and eventually converges to the asymptotic 
value predicted by the dilute instanton gas model at higher temperature \cite{Bonati:2015vqz,Bonati:2015sqt,GrillidiCortona:2015jxo}, namely 
\begin{equation}
b_2^{\text{inst}}=-\frac{1}{12} \simeq -0.083.
\end{equation}     
The upper panel shows that for $\mu=300~\text{MeV}$, $b_2$ drops significantly at the critical point $(T,\mu)=(30,300)~\text{MeV}$,  
and then increases with $T$ toward $b_2^{\text{inst}}$ at high temperature. The bottom panel of Fig. \ref{fig:b2QM} displays $b_2$ 
as the function of $\mu$ calculated in QM for different temperatures. We notice that for all the cases, $b_2$ also approaches $b_2^{\text{inst}}$
at large $\mu$. For a fixed higher temperature, $b_2$ decreases monotonically with $\mu$. For a fixed lower temperature, an inverted peak 
appears at the chiral transition point and the deepest one corresponds to the critical end point. This mean that the deepest peak of $b_2$ 
can be used to indicate the QCD critical point. Such a conclusion is consistent with that obtained in the NJL model \cite{Gong:2024cwc}. 
In addition, the calculation based on the PQM model gives the similar results, as displayed in Fig. \ref{fig:b2PQM}.              
       
\begin{figure}[!t]
\centering
\begin{minipage}[h]{0.45\textwidth}
\centering
\includegraphics[width=1.0\textwidth]{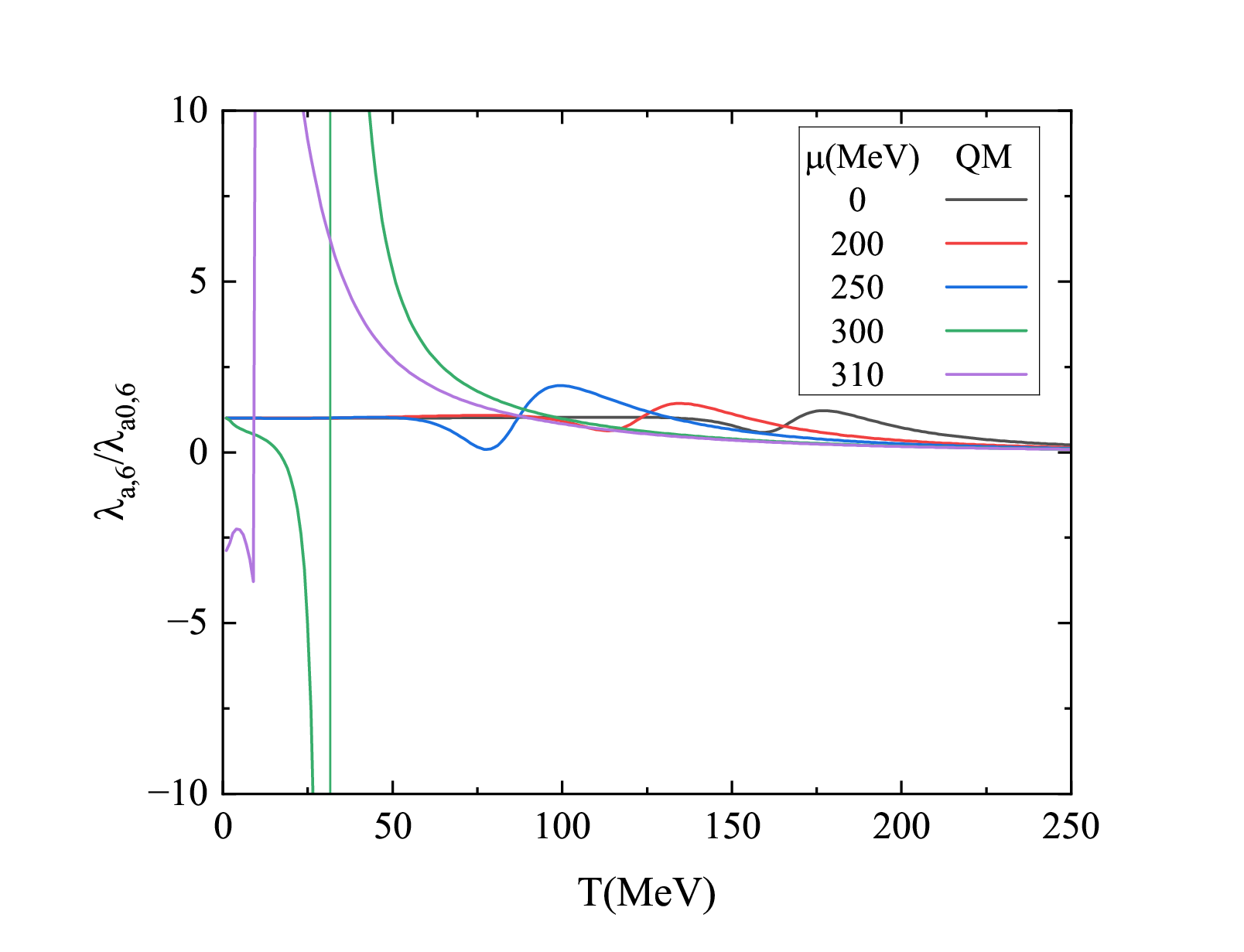}
\end{minipage}
\begin{minipage}[h]{0.45\textwidth}
\centering
\includegraphics[width=1.0\textwidth]{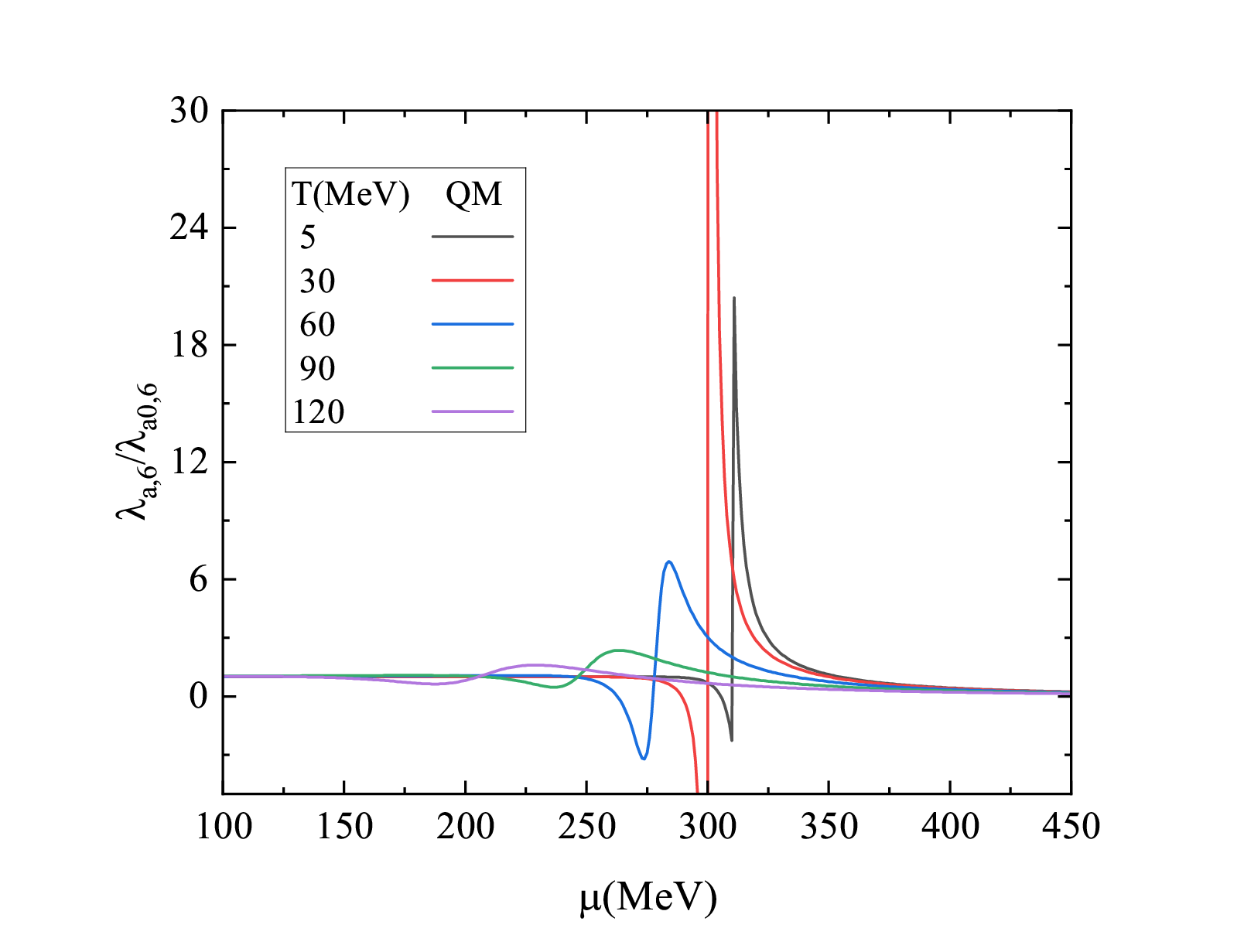}
\end{minipage}
\caption{The normalized axion sextic self-coupling constant $\lambda_{a,6}$ as the function of temperature at different quark chemical potentials (upper panel) 
and as the function of quark chemical potential at different temperatures (lower panel), obtained in the QM model. $\lambda_{a0,6}$ is the vacuum value of $\lambda_{a,6}$ 
calculated in QM.}
\label{fig:sc6QM}
\end{figure}

\begin{figure}[!t]
\centering
\begin{minipage}[h]{0.45\textwidth}
\centering
\includegraphics[width=1.0\textwidth]{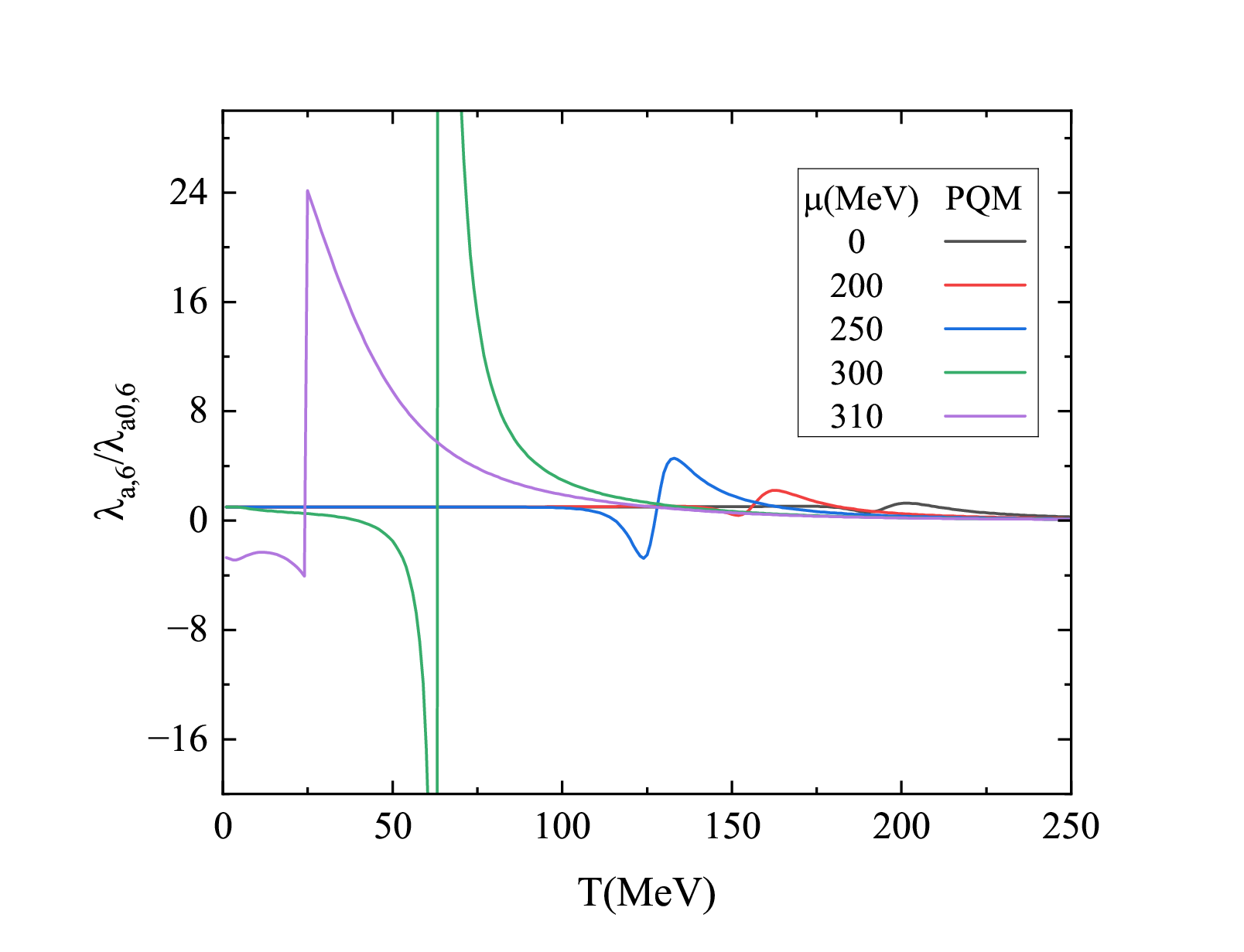}
\end{minipage}
\begin{minipage}[h]{0.45\textwidth}
\centering 
\includegraphics[width=1.0\textwidth]{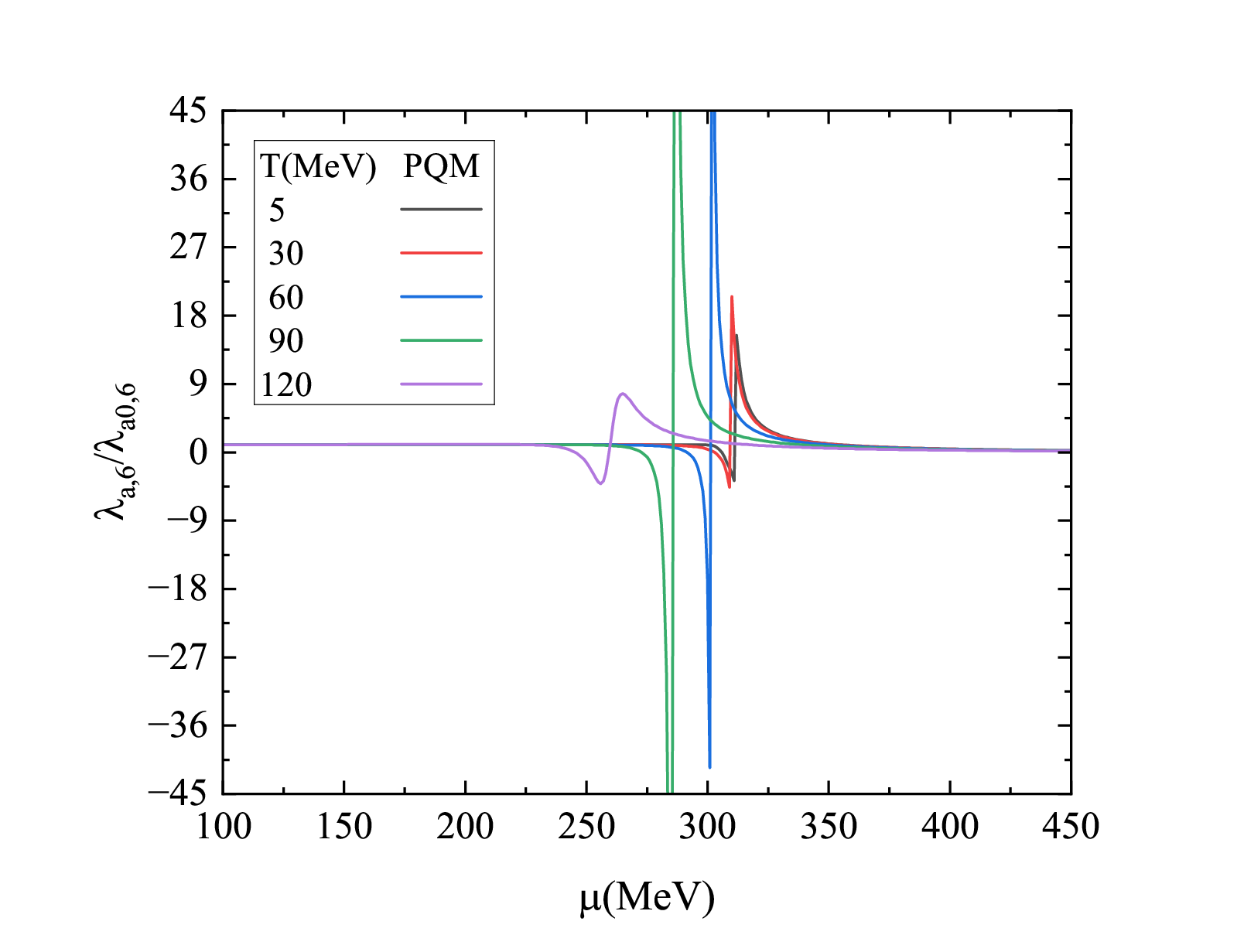}
\end{minipage}
\caption{Same as Fig. \ref{fig:sc6QM}, but obtained within the PQM model.}
\label{fig:sc6PQM}
\end{figure}

Figures \ref{fig:cp4QM} and \ref{fig:cp4PQM} display the quartic self-coupling $\lambda_{4,a}$ of QCD axion vs the temperature and 
chemical potential obtained in QM and PQM, respectively. Both models show that, compared to its vacuum value, the coupling enhances significantly 
at and near the critical point and reduces obviously in the chiral symmetric phase far away from the critical point, which is qualitatively 
consistent with the previous studies bases on the NJL model \cite{Lu:2018ukl,Gong:2024cwc}.

\begin{table}[htbp]
\centering
\begin{tabular}{cccc}
\hline
$m_{\sigma}(\mathrm{MeV})$ & $b_2$ & $b_4$ & $\lambda_a^6\times f_a^6$ \\
\hline
$400$ & $-0.0252844$ & $0.000423698$ & $(13.4491$MeV$)^6$ \\
$500$ & $-0.022461$ & $0.000255862$ & $(12.3648$MeV$)^6$  \\
$600$ & $-0.0209273$ & $0.000189485$ & $(11.7611$MeV$)^6$ \\
$700$ & $-0.0200026$ & $0.000153791$ & $(11.359$MeV$)^6$ \\
$800$ & $-0.0194023$ & $0.00013141$ & $(11.0651$MeV$)^6$ \\
$900$ & $-0.0189908$ & $0.000116148$ & $(10.8398$MeV$)^6$ \\
$1000$ & $-0.0186965$ & $0.000105179$ & $(10.662$MeV$)^6$ \\
\hline
\end{tabular}
\caption{Cumulants $b_2$,$b_4$ and the sixth self-coupling $\lambda_{a,6}$ in vacuum predicted by QM for 
different $m_\sigma$.}
\label{222}
\end{table}

The variation of the normalized sixth-order cumulant $b_4$ with temperature at different chemical potentials is shown in Fig. \ref{fig:b4QM}.
At vacuum, the (P)QM model with $m_{\sigma} = 500~\text{MeV}$ gives 
\[
b_4^{\mathrm{(P)QM}} =0.000256,
\]
which is close to but larger than the ChPT result $b_4=0.00017(6)$ in the isospin limit. The dependence of the vacuum $b_4$ on the value of $m_{\sigma}$ 
in QM with other model parameters unchanged is listed in Table \ref{222}. We find that $b_4$  decreases monotonically with $m_{\sigma}$,  and the listed $b_4$ 
values basically lie within the range  $\left|b_4\right| \lesssim 4 \times 10^{-4}$ predicted by the lattice SU(3) pure gauge theory \cite{Bonati:2015sqt}, except for $m_{\sigma}=400 \text{MeV}$. 
The upper panel shows that at high temperatures, $b_4$ always approaches the value $b_4^{\mathrm{inst}} = \frac{1}{360} \approx 0.00278$ given by the dilute 
instanton gas model. For zero and small chemical potentials $\mu=0, 100, 200 ~\text{MeV}$, $b_4$ remains almost constant at low temperatures and then rises rapidly 
with T near the crossover temperature. For the larger chemical potential $\mu=250~\text{MeV}$, $b_4$ first decrease with $T$ near and before the pseudo-critical 
temperature, and then increases swiftly with $T$ across the crossover region. For the even larger $\mu=310~\text{MeV}$, $b_4$ changes sign at the critical 
temperature $T_c$ of the first order transition: it is negative below $T_c$ and positive above $T_c$, and grows significantly in the chiral symmetric side.     
We see that $b_4$ changes sign and becomes divergent at the critical endpoint: it approaches negative infinity on the left side of the critical point 
and positive infinite on the right side. This is quite different from the critical behavior of the cumulant $b_2$, which means the Taylor expansion of 
$F(\theta,T,\mu)$ fails in the vicinity of the critical point. The lower panel shows the dependence of $b_4$ as a function of the chemical potential for 
different temperatures in QM. We see the similar critical thermal behavior at and near the critical point and the cumulant $b_4$ also converges to the 
uniform value $b_4^{\mathrm{inst}}$ at high chemical potentials. This later implies that $b_4$ remains largely insensitive to hot and dense media. The 
peculiar critical behavior of $b_4$ can be used to indicate the critical point of QCD. All the above features are also observed in Fig. \ref{fig:b4PQM} 
within the PQM formalism.

\begin{figure}[!t]
\centering
\begin{minipage}[h]{0.45\textwidth}
\centering
\includegraphics[width=1.0\textwidth]{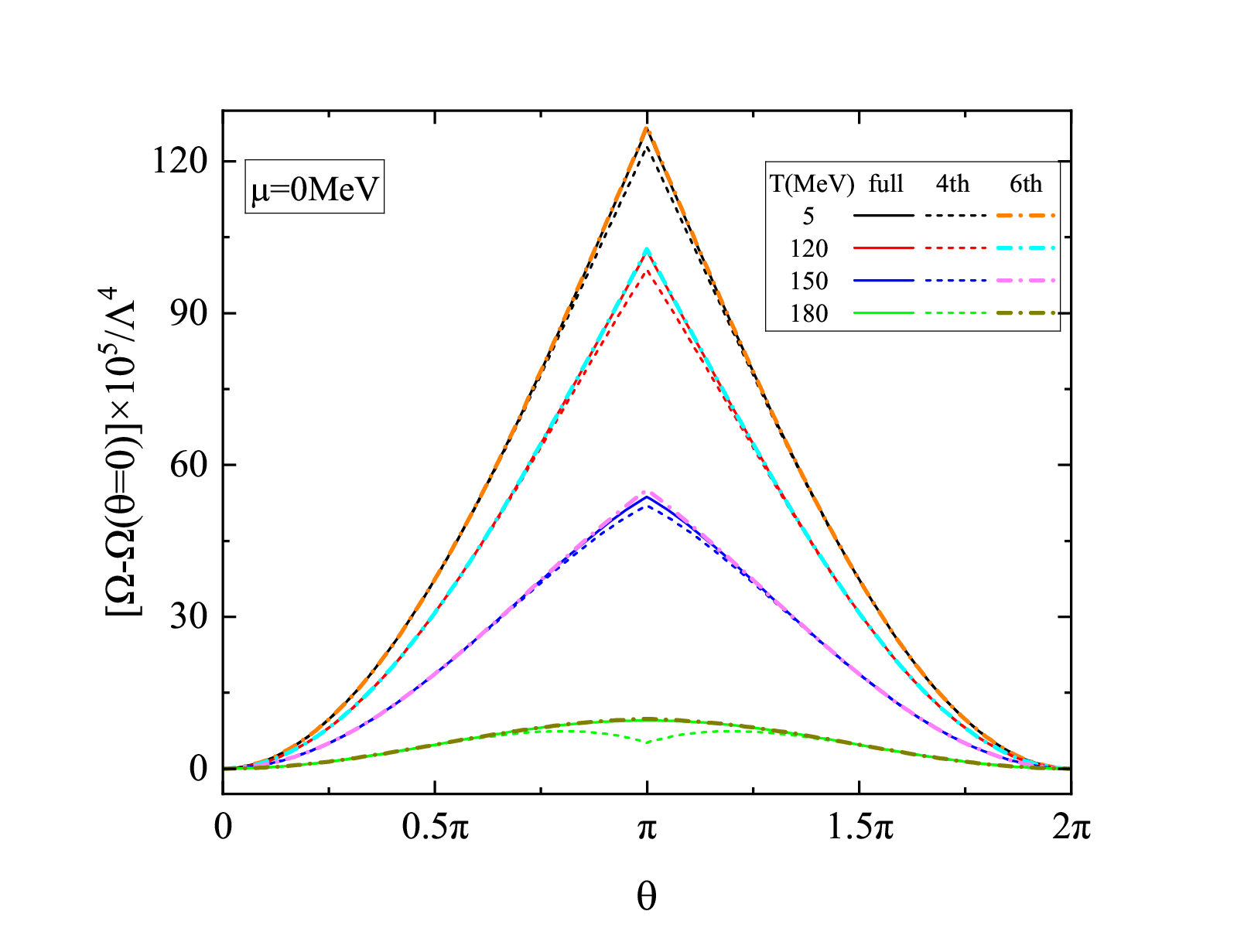}
\end{minipage}
\begin{minipage}[h]{0.45\textwidth}
\centering
\includegraphics[width=1.0\textwidth]{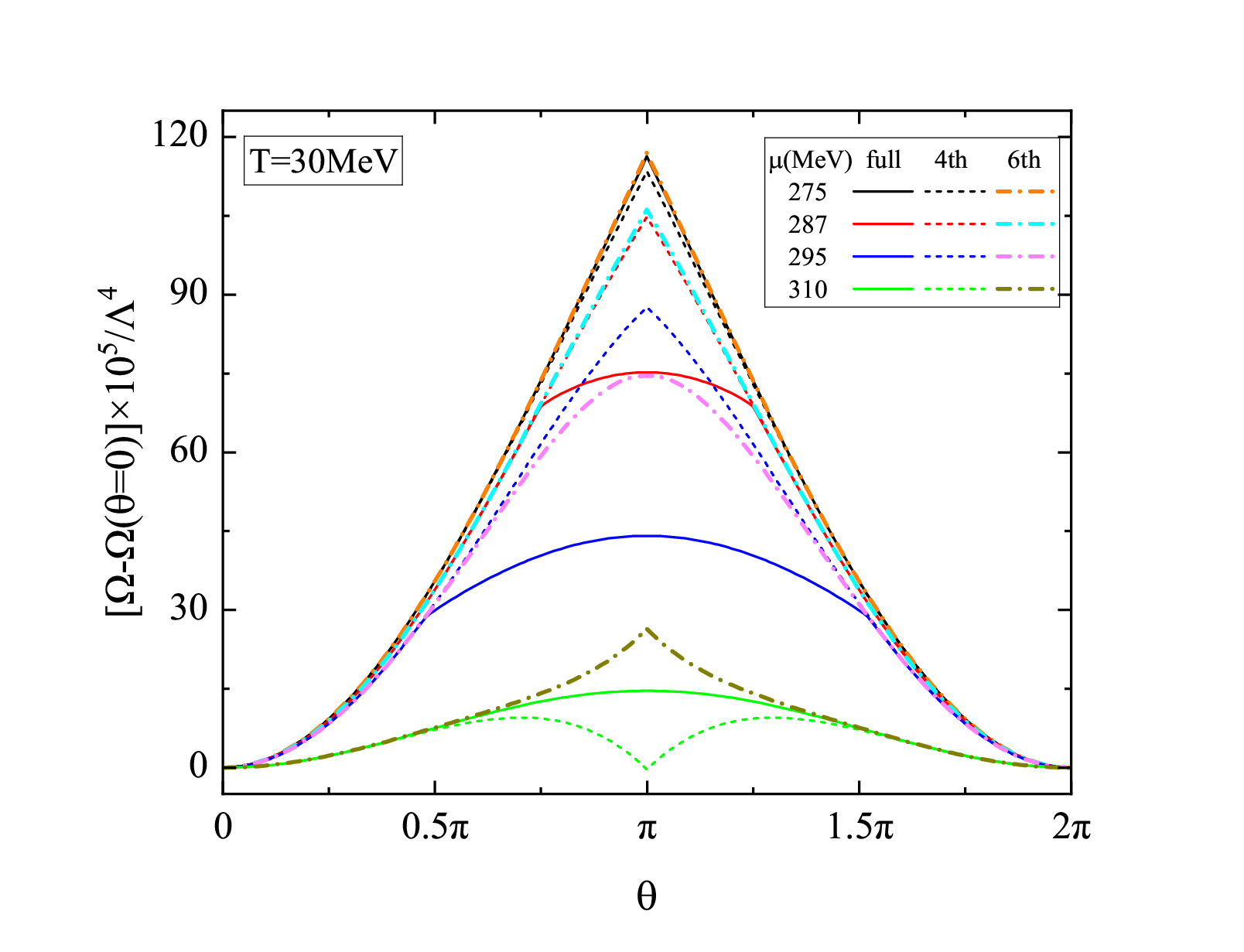}
\end{minipage}
\caption{Full potential $\Omega(\theta)$ and its fourth- and sixth-order Taylor expansions within the QM formalism: upper panel for $\mu=0$ 
at different temperatures and lower panel for $T=30~\text{MeV}$  at different chemical potentials.   }
\label{fig:taylorQM}

\end{figure}
\begin{figure}[!t]
\centering
\begin{minipage}[h]{0.45\textwidth}
\centering
\includegraphics[width=1.0\textwidth]{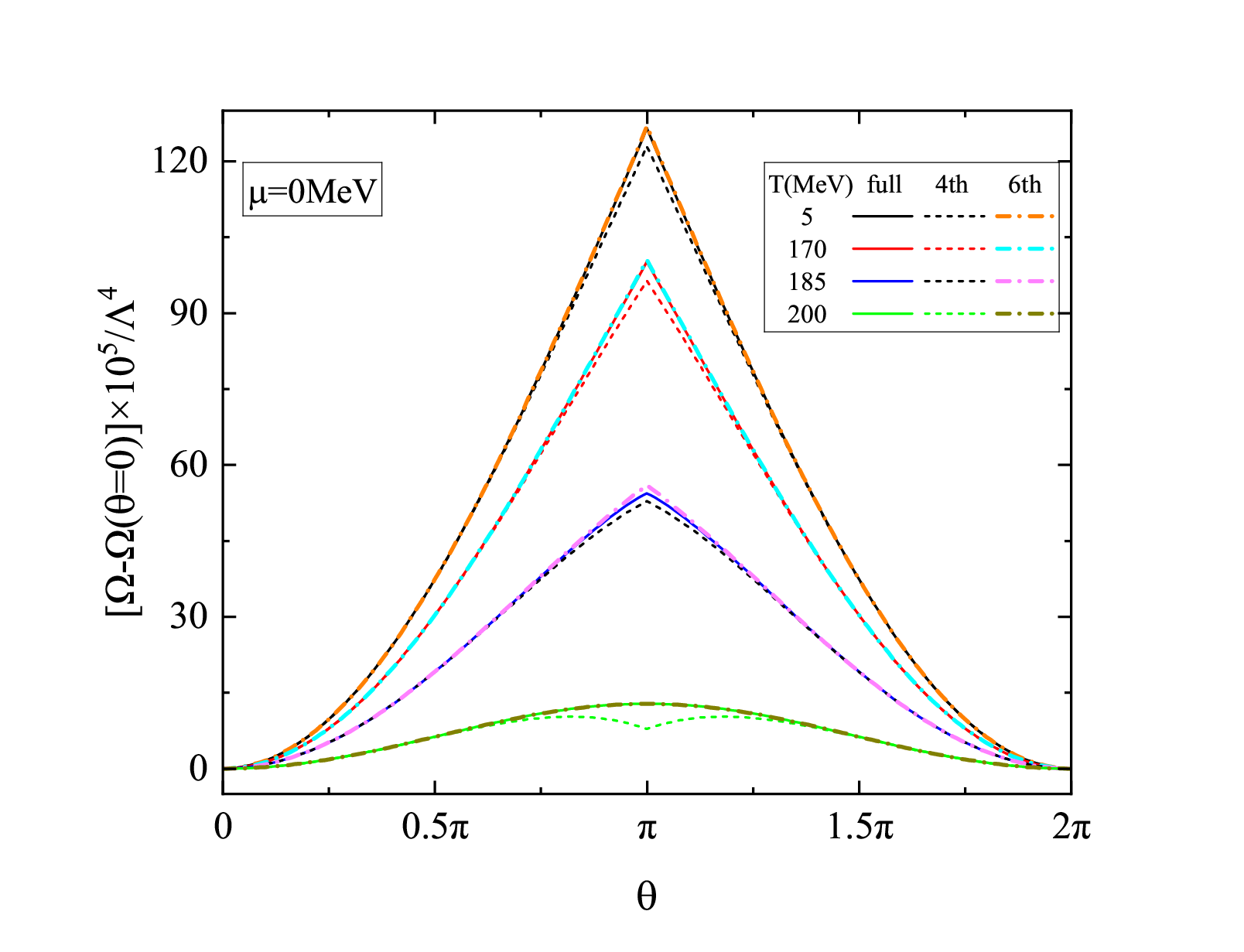}
\end{minipage}
\begin{minipage}[h]{0.45\textwidth}
\centering
\includegraphics[width=1.0\textwidth]{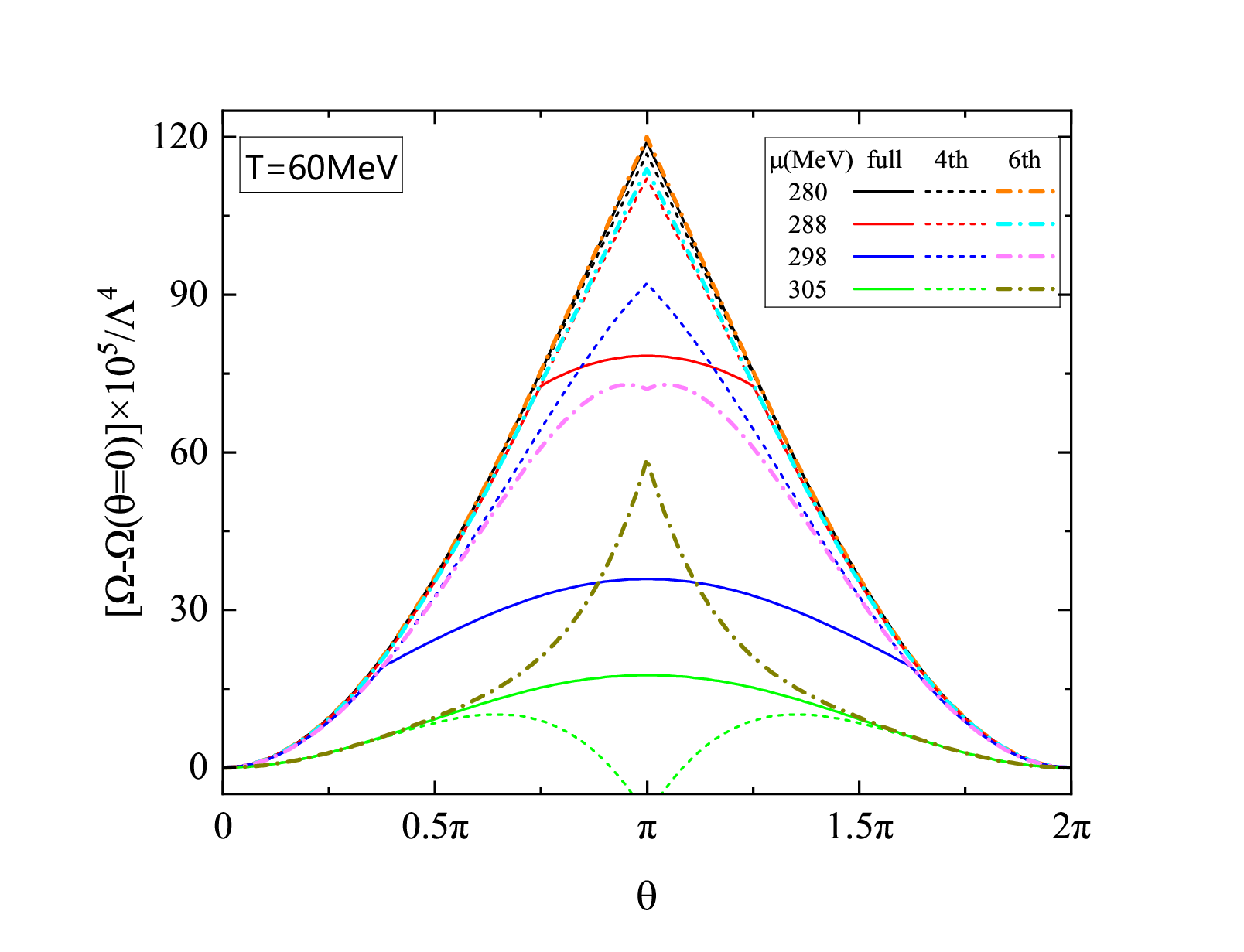}
\end{minipage}
\caption {Full potential $\Omega(\theta)$ and its fourth- and sixth-order Taylor expansions within the PQM formalism: upper panel for $\mu$=0 at different temperatures and lower panel for $T=60~\text{MeV}$  at different chemical potentials.}
\label{fig:taylorPQM}
\end{figure}

The normalized sixth-order self-coupling constant $\lambda_{a,6}$ of QCD axion as the function of temperature and quark chemical potential within the QM model 
and PQM model is shown in Figs. \ref{fig:sc6QM} and \ref{fig:sc6PQM}, respectively. The curves for $\lambda_{a,6}$ in these two figures are obtained from $b_4$ 
displayed in Figs. \ref{fig:b4QM} and \ref{fig:b4PQM}, respectively, via Eq. \eqref{eq:sc6}. As expected, all the figures show that 
$\lambda_{a,6}$ also changes sign across the phase boundary near the critical point and becomes divergent at this point: the sixth self-coupling turns 
attractive in the chiral symmetry broken phase close to the critical point and remain repulsive in the chiral restored phase in the vicinity. In particular, the 
self-coupling $\lambda_{a,6}$ may be very large near the critical point, which may significantly alter the physics of axion stars if they formed during the 
QCD phase transition in the early University or changes the axion condensation in Neutron Stars.            

\subsection{Range of accuracy of the truncated Taylor expansion of the thermodynamical potential as the function of $\theta$}

Unlike the lattice QCD simulations, we can calculate the full potential $\Omega(\theta)$ at the mean field level in the (P)QM model. It is meaningful 
to assess the Taylor expansion approximation of $\Omega(\theta)$ in the formalism of effective theory if the topological susceptibility and cumulants 
$b_2$ and $b_4$ are known.

Figure. \ref{fig:taylorQM} show the full potential $\Omega(\theta)$ and its fourth- and sixth-order Taylor expansions in the range $0 \leq\theta\leq 2\pi$ at 
finite temperature and chemical potential within the QM formalism \footnote{For comparison with the NJL results, the potential 
is uniformly rescaled by the fourth power of $\Lambda=590 \text{MeV}$, the cutoff parameter used in the NJL model.}. The left dashed and dash-dotted 
lines in the figure correspond to the Taylor expansions centered at \(\theta=0\) using \(\theta\) as the variable, while the right ones in the figure 
correspond to the Taylor expansions centered at \(\theta=2\pi\) using \(\theta-2\pi\) as the variable. Since the curves in the figure are symmetric 
left and right, we only compare the left ones in the interval $0 \leq\theta\leq \pi$.  

The upper panel demonstrate the mean field potential and the fourth- and sixth-order Taylor approximations of $\Omega(\theta)$ vs the $\theta$ angle at 
zero chemical potential for different temperatures. The cusps appearing on the curves at $T=5, 120, 150~\text{MeV}$ indicate the Danshen's phenomenon, where 
first-order transitions occur at $\theta=\pi$ and the condensate $\eta$ flips its sign. By contrast, No cusp is observed on the curve for the higher temperature 
$T=180~\text{MeV}$, as the condensate $\eta$ is suppressed to zero at $\theta=\pi$. We see that for all cases, the curve of the full potential $\Omega(\theta)$ 
almost coincident with that of the sixth-order Taylor approximation over the entire range $0\leq\theta \leq \pi$. In contrast, the fourth-order Taylor 
expansion also matches quite well with the full potential $\Omega(\theta)$ for $0\leq\theta \lesssim 0.65\pi$ and deviates approximately in the 
interval $0.65 \pi \lesssim \theta \leq \pi$. All curves show that the deviation increases with $\theta$ toward $\pi$, yet it remains fairly small 
even at $\theta=\pi$, except for $T=180 \text{MeV}$, where both condensates $\sigma$ and $\eta$ are substantially suppressed. 

The lower panel illustrates the mean field potential and the fourth- and sixth-order Taylor approximations of $\Omega(\theta)$ vs the $\theta$ angle 
at a low temperature $T=30~\text{MeV}$ for four different chemical potentials around $\mu=300~\text{MeV}$. These $(\mu,T)$ points are close to the 
critical point $(\mu=300, T=30)~\text{MeV}$ at zero $\theta$ and we will check the range of accuracy of the Taylor expansions near the phase boundary
around the critical point. For $\mu=275~\text{MeV}$, the first-order transition still emerges at $\theta=\pi$, which is similar to the three lower temperature curves in 
the upper panel. In this case, the sixth-order Taylor expansion also agrees excellently with the full potential, while the fourth-order approximation 
only deviates slightly for large $\theta$ near $\pi$. For $\mu=287$ and $295~\text{MeV}$, the first-order phase transitions occur at the critical 
angle $\theta_{c1}=0.75 \pi$ and  $\theta_{c2}=0.5 \pi$, respectively, as indicated by the new cusps on the curves of the full potentials. We see 
that both the fourth- and sixth-order Taylor expansions deviate significantly from the full potential for $\theta_{c1}\leq\theta\leq \pi$ and 
$\theta_{c2}\leq\theta\leq \pi$, even if they are nearly perfectly coincident with the full potential in the intervals $0 < \theta < \theta_{c1}$ 
and $0 < \theta < \theta_{c2}$. In other words, the fourth- or sixth-order Taylor approximations work well in the chiral breaking phase in the 
range $\theta < \theta_c$,  but fail in the chiral restored phase for $\theta>\theta_c$, where $\theta_c$ is the critical value of $\theta$ for 
the first-order transition. For $\mu=310~\text{MeV}$, both condensates $\sigma$ and $\eta$ becomes very small and there is no phase transition in 
the interval $0 \leq \theta < 2\pi$. We see that both Taylor expansions only work well in the small $\theta$ range $0 \leq \theta < 0.5-0.6 \pi$ 
and the sixth-order approximation only improves slightly compared to the fourth-order expansion. This means that in the chiral symmetric phase, much 
higher order cumulants are needed in the potential Taylor expansion for large $\theta$, even no first-order phase transition occur.       
 
The calculations in PQM yield the similar results, as displayed in Fig. \ref{fig:taylorQM}. We thus conclude that the sixth-order or even 
the fourth-order Taylor expansion of the potential with respect to $\theta$ provide a fairly good approximation in the chiral symmetry breaking 
phase at finite temperature and density. When a first-order phase transition happens, the Taylor expansion fails in the chiral symmetric phase 
with large $\theta$. If the whole range of $0\leq\theta \leq \pi$ is in the chiral symmetric phase, the Taylor expansion up to sixth-order
is quite well at small density and high temperature, but becomes bad at low temperature and relatively large density.  
We can further infer that the fourth- or sixth-order Taylor expansions of the potential as the function of $\theta$ in lattice QCD simulations 
may converge well up to large $\theta=\pi$, provided the topological susceptibility and the cumulants $b_2$ and $b_4$ are calculated exactly 
at finite temperature where no first-order transition takes place.

\subsection{Comparison with the (P)NJL results}        
 
The fourth-order cumulant $b_4$ and the range of accuracy of the Taylor expansion have also been investigated within the NJL model with and without 
considering the Polyakov loop dynamics. Unlike the QM with $m_{\sigma}=500 \text{MeV}$, the NJL-type model usually predicts a too large chiral 
pseudo-critical temperature at zero density comparing to the lattice QCD simulations. Adopting the same model used in Ref. \cite{Lu:2018ukl}, 
we find that the vacuum $b_4$ takes the value
\[
b_4^{\mathrm{(P)NJL}} = 0.000129754,
\]
which is about $27\%$ less than the ChPT prediction. The thermal behaviors of $b_4$ and $\lambda_{a,6}$ obtained in (P)NJL are qualitatively 
consistent with that obtained in (P)QM, and more details are given in the Appendix.  

\section{ \label{sect:conclusion} Conclusion and Outlook }

In this study, the lower-order cumulants of the topological charge distribution are investigated in the (P)QM model, which has the merit 
of renormalization. We focus on the sixth-order cumulant $b_4$ at finite temperature and density, which is directly connected to the sixth-order 
self-coupling $\lambda_{a,6}$ of QCD axion. The medium properties of $b_4$ or $\lambda_{a,6}$ are hard to evaluate within lattice QCD simulations. 
Furthermore, the range of accuracy of the Taylor expansion of the thermal dynamic potential as a function of $\theta$ has been systematically analyzed 
by comparing to the mean field full potential obtained within the (P)QM formalism.

We find that under the isospin symmetric condition, the vacuum $b_4$ in QM (also in NJL) is close to the ChPt result and lies within the bound 
provided by lattice estimates, $\left|b_4\right| \lesssim 4 \times 10^{-4}$. It remains almost constant in the chiral symmetry breaking phase at 
low temperature and/or chemical potential, and approaches to the prediction by the dilute instanton gas model at high temperature and/or 
chemical potential. Near the chiral phase boundary, $b_4$ flips sign and exhibits non-monotonicity. In particular, it becomes divergent at the 
QCD critical point, which is quite different from the critical behavior of the fourth-order cumulant $b_2$. Correspondingly, the self-coupling $\lambda_{a,6}$ 
of QCD axion also displays a nontrivial behavior across the phase boundary, which first deceases to a negative value on the chiral symmetry 
breaking side, and then increases significantly and changes sign on the chiral restored side. Especially, $\lambda_{a,6}$ changes from a negative 
infinity to the positive infinity across the critical point. Such pronounced non-monotonic behavior near the phase boundary and the divergence at 
the critical point may significantly influence the possible formation of an axion BEC inside a proto-neutron star, whose thermal evolution could 
sweep through the QCD critical point.

The range of accuracy analysis reveals that the sixth-order or even the fourth-order Taylor expansion of the potential works quite well if the whole 
range of $0\leq\theta <\pi$ belongs to the hadronic phase for a fixed $T$-$\mu$ point. When a first-order phase transition occurs at $\theta=\theta_c$ 
for a fixed $T$-$\mu$ point, the range of accuracy of the fourth- and sixth-order Taylor expansions shrink to $\theta_c$, as expected. When 
the chiral restored phase remains in the interval $0\leq\theta <\pi$, the sixth-order approximation of the potential is still quite good for vanishing 
density, but fails for low temperature and high chemical potential. In the latter case, higher-order cumulants are required for the Taylor expansion
approximation. 

In this paper, the isospin breaking effect due to the mass difference of u and d quarks is missed, which may yield a negative $b_4$ in 
vacuum. In compact stars, charge neutrality and $\beta$-equilibrium constraints give rise to density mismatch between u and d quarks.  
It is interesting to investigate how these isospin asymmetries affect the medium properties of $b_4$ and $\lambda_{a,6}$, especially in the vicinity 
of the critical point. For the more realistic case, the strange quark degree of freedom should be included, and the present study can be extended
to the $2$+$1$ flavor system \cite{Schaefer:2008hk,Jiang:2015xqz}. We will leave these topics in our future publications.     

\vspace{5pt}
\noindent{\textbf{\large{Acknowledgements}}}

This work was supported by the National Natural Science Foundation of China (NSFC) under Grant No. 11875127.

\appendix \label{sect:appendix}

\begin{figure}[!htbp]
\centering
\begin{minipage}[h]{0.45\textwidth}
\centering
\includegraphics[width=1.0\textwidth]{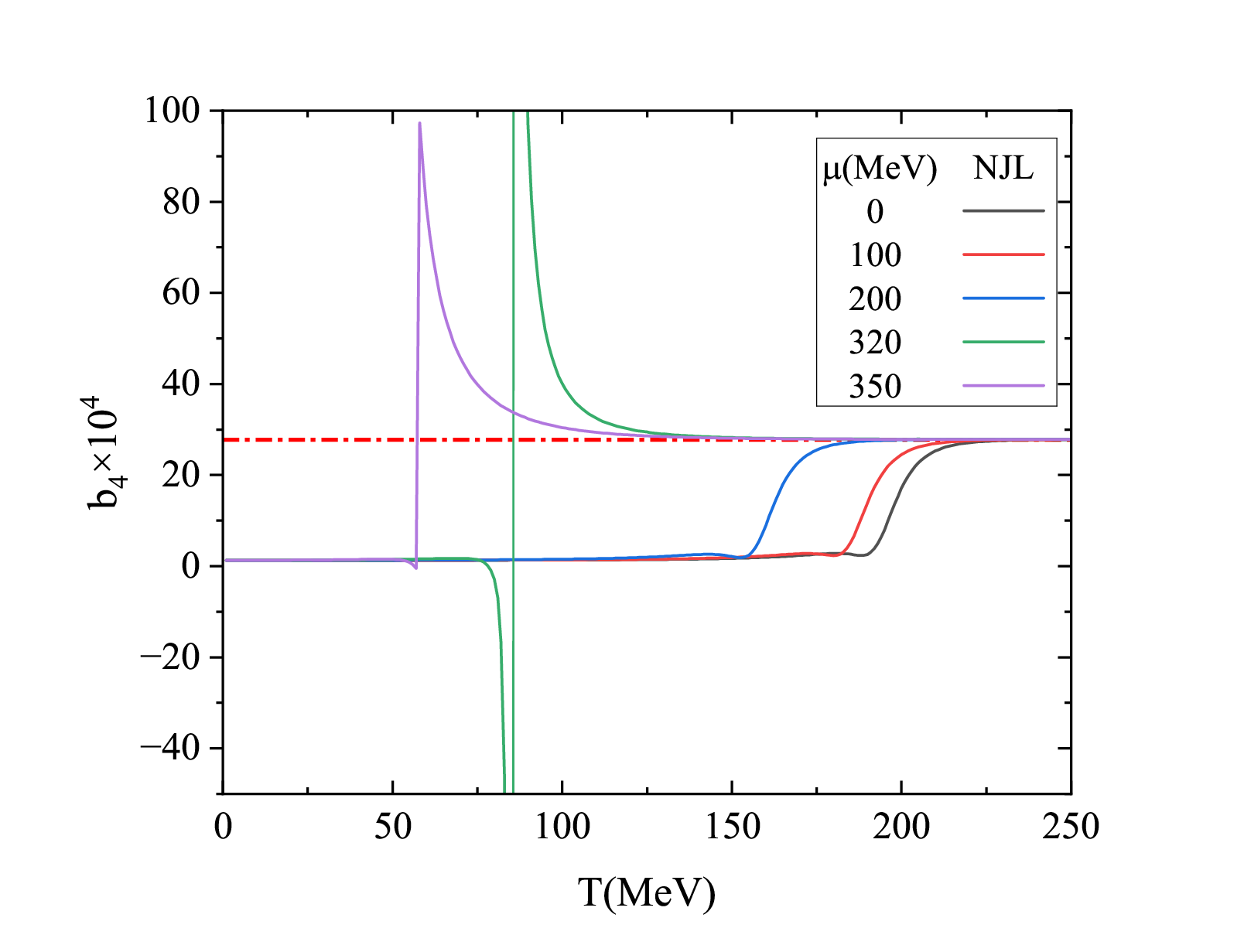}
\end{minipage}
\begin{minipage}[h]{0.45\textwidth}
\centering
\includegraphics[width=1.0\textwidth]{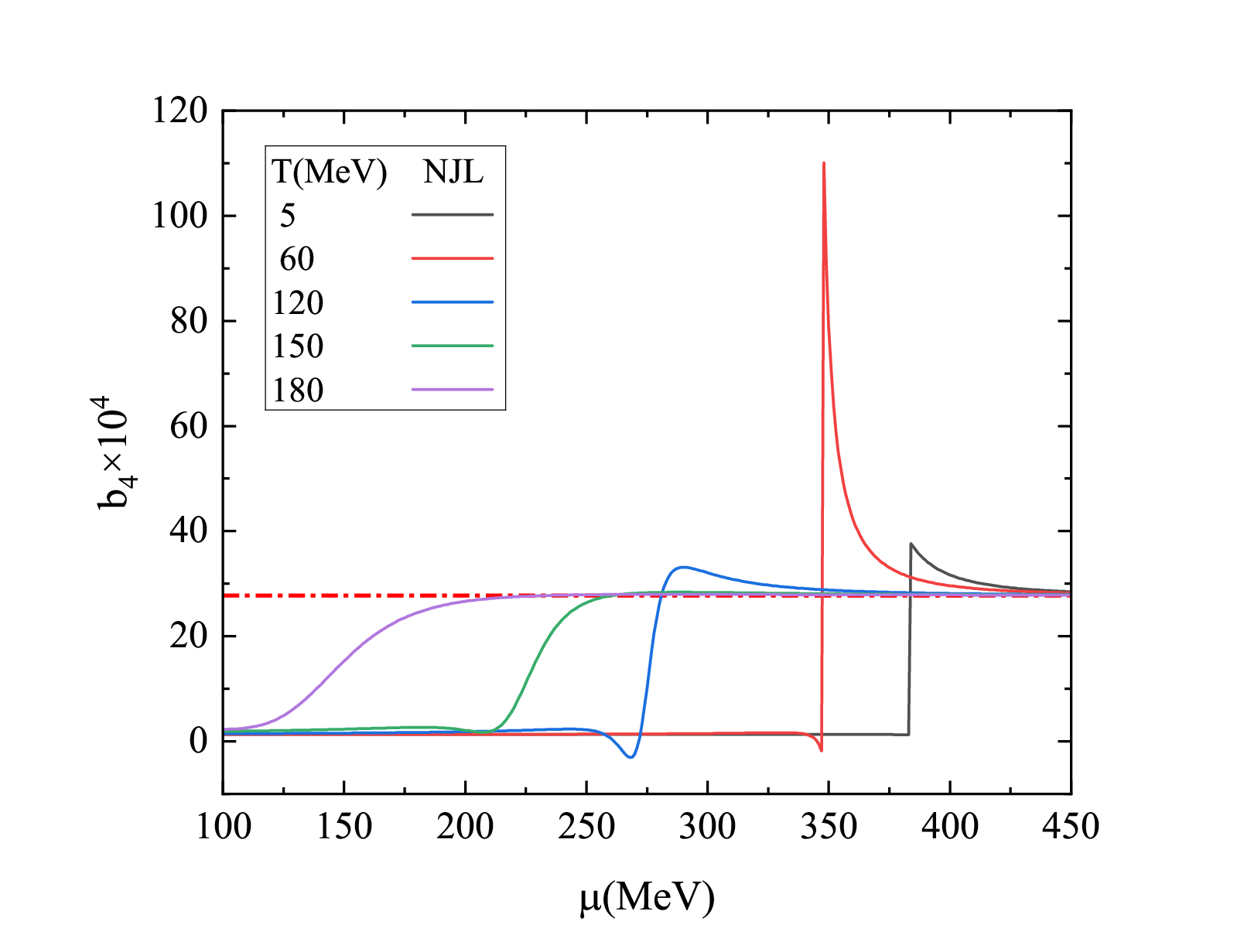}
\end{minipage}
\caption{The normalized sixth-order cumulant $b_4$ of the QCD topological charge distribution as the function of temperature at different quark chemical potentials (upper panel) 
and as the function of quark chemical potential at different temperatures (lower panel), obtained in the NJL model. The red dash-dotted line is the same as that in Fig. \ref{fig:b4QM}. }
\label{fig:b4NJL}
\end{figure}

\begin{figure}[!htbp]
\centering
\begin{minipage}[h]{0.45\textwidth}
\centering
\includegraphics[width=1.0\textwidth]{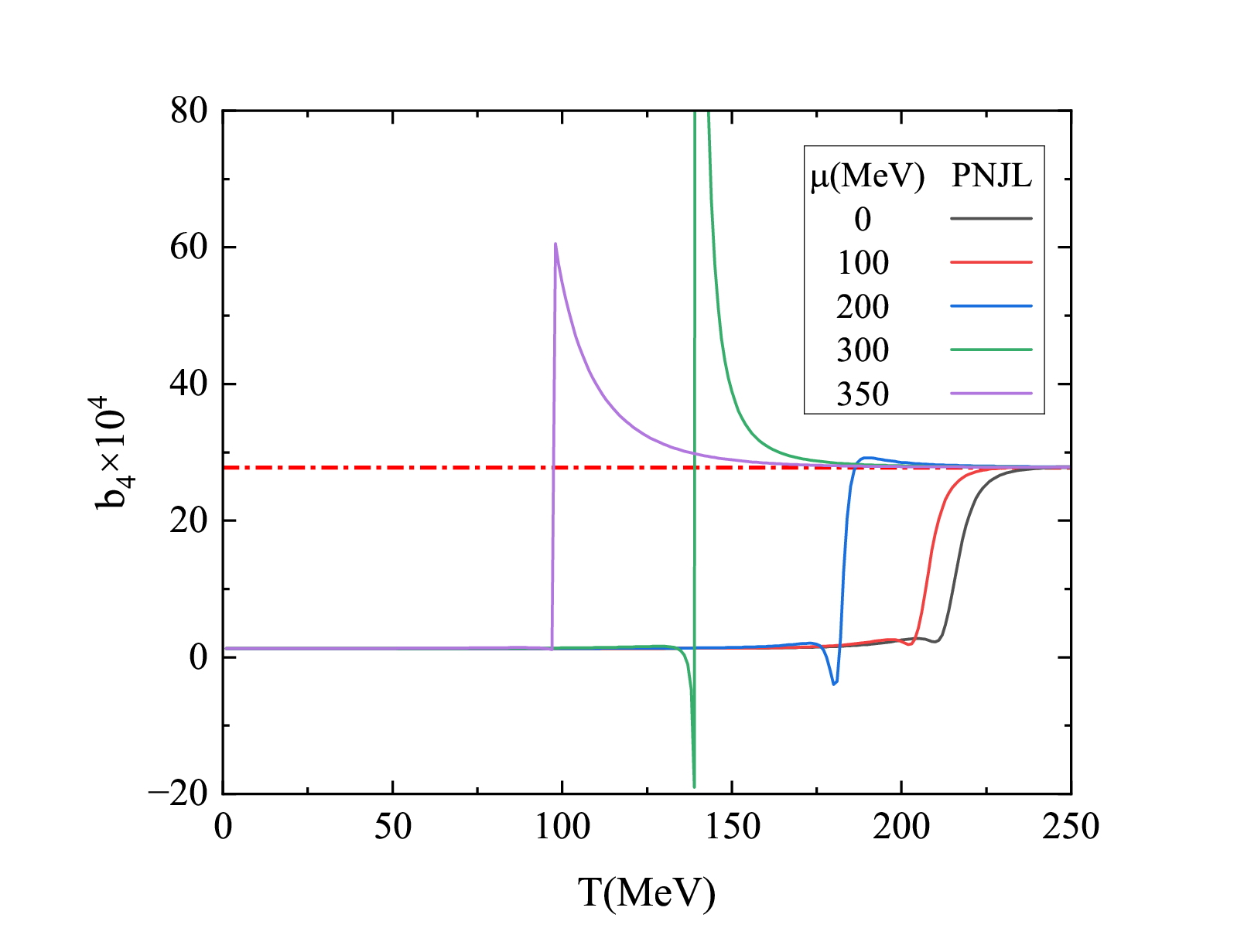}
\end{minipage}
\begin{minipage}[h]{0.45\textwidth}
\centering
\includegraphics[width=1.0\textwidth]{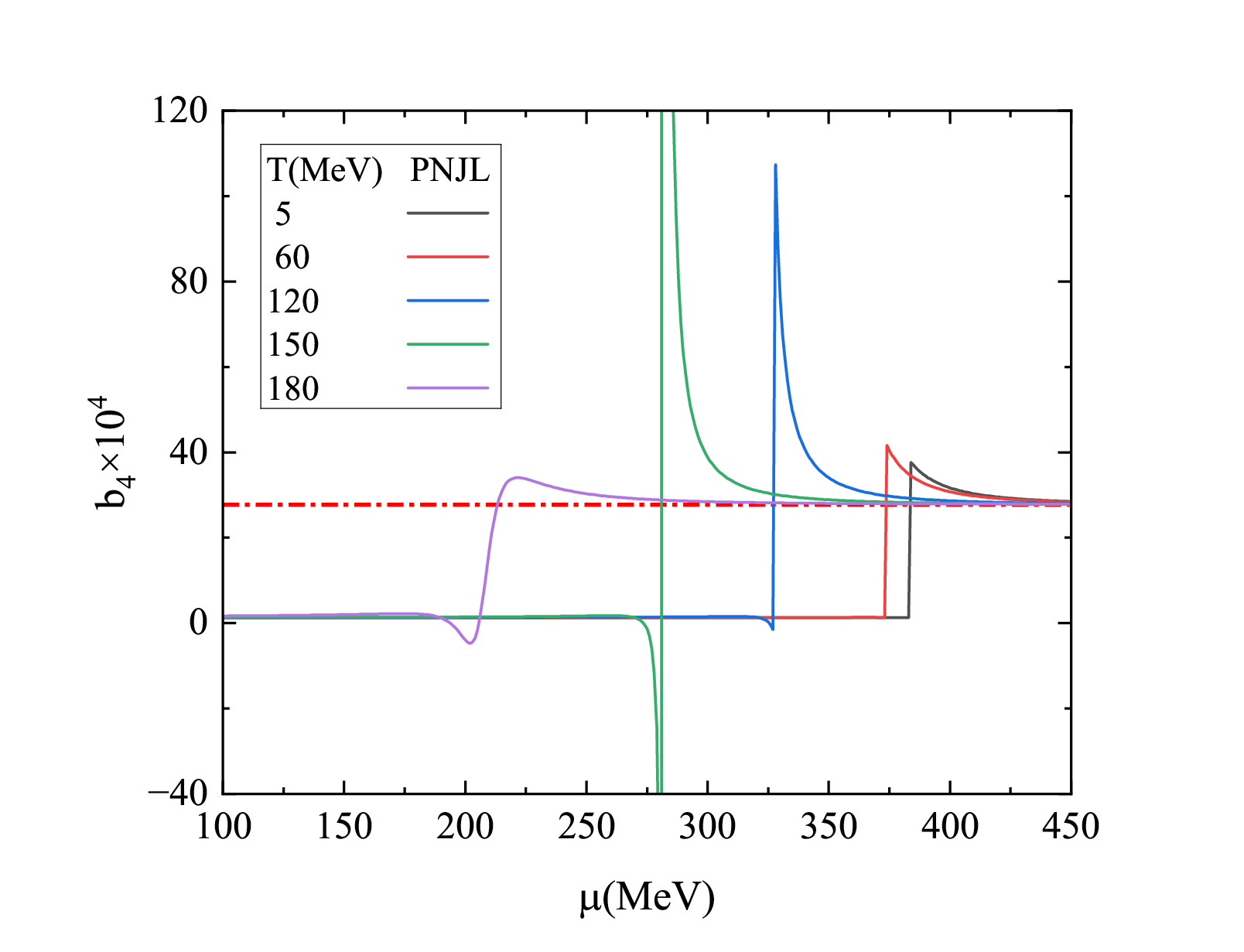}
\end{minipage}
\caption{ Same as Fig. \ref{fig:b4NJL}, but obtained in the PNJL model.}
\label{fig:b4PNJL}
\end{figure}

\begin{figure}[!htbp]
\centering
\begin{minipage}[h]{0.45\textwidth}
\centering
\includegraphics[width=1.0\textwidth]{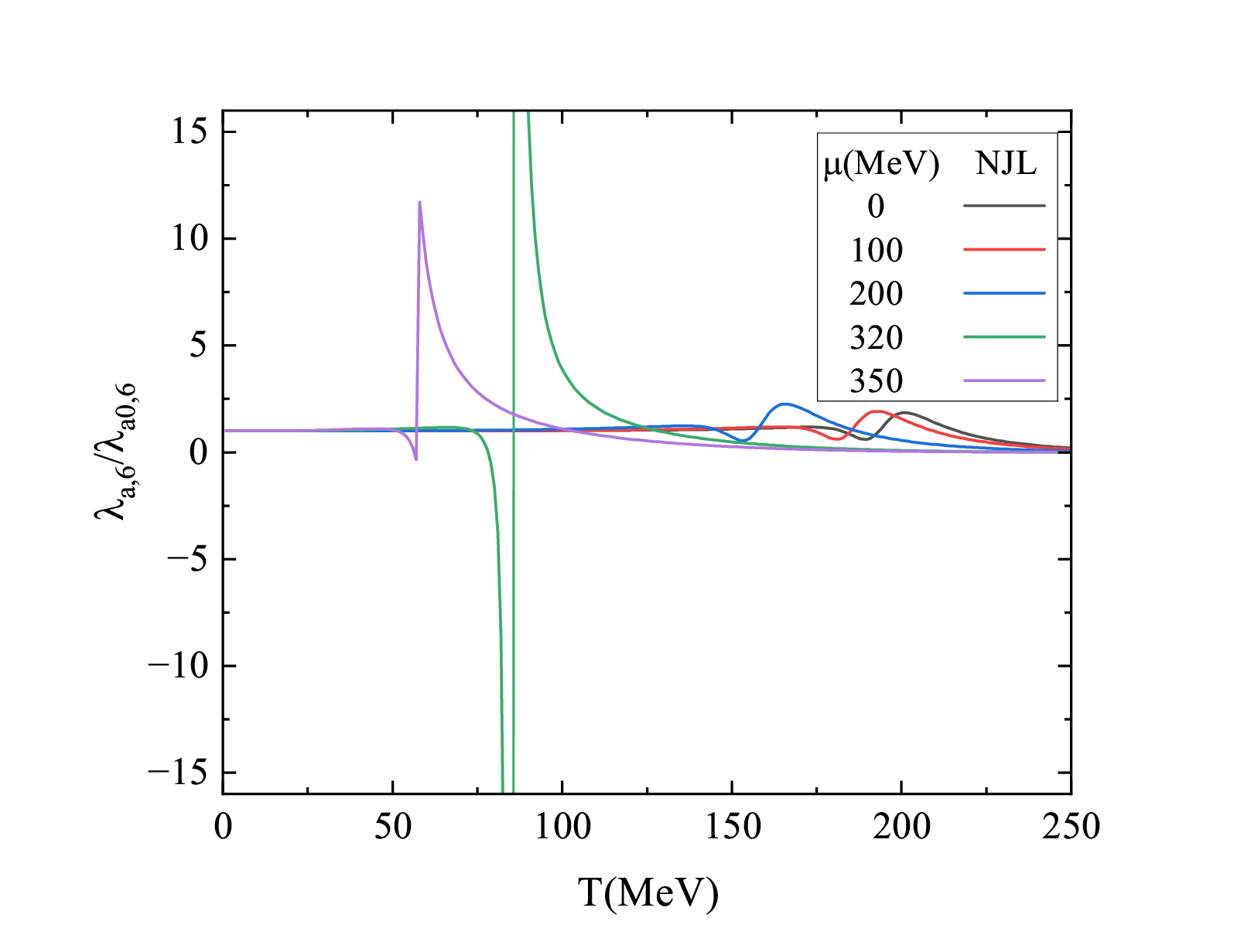}
\end{minipage}
\begin{minipage}[h]{0.45\textwidth}
\centering
\includegraphics[width=1.0\textwidth]{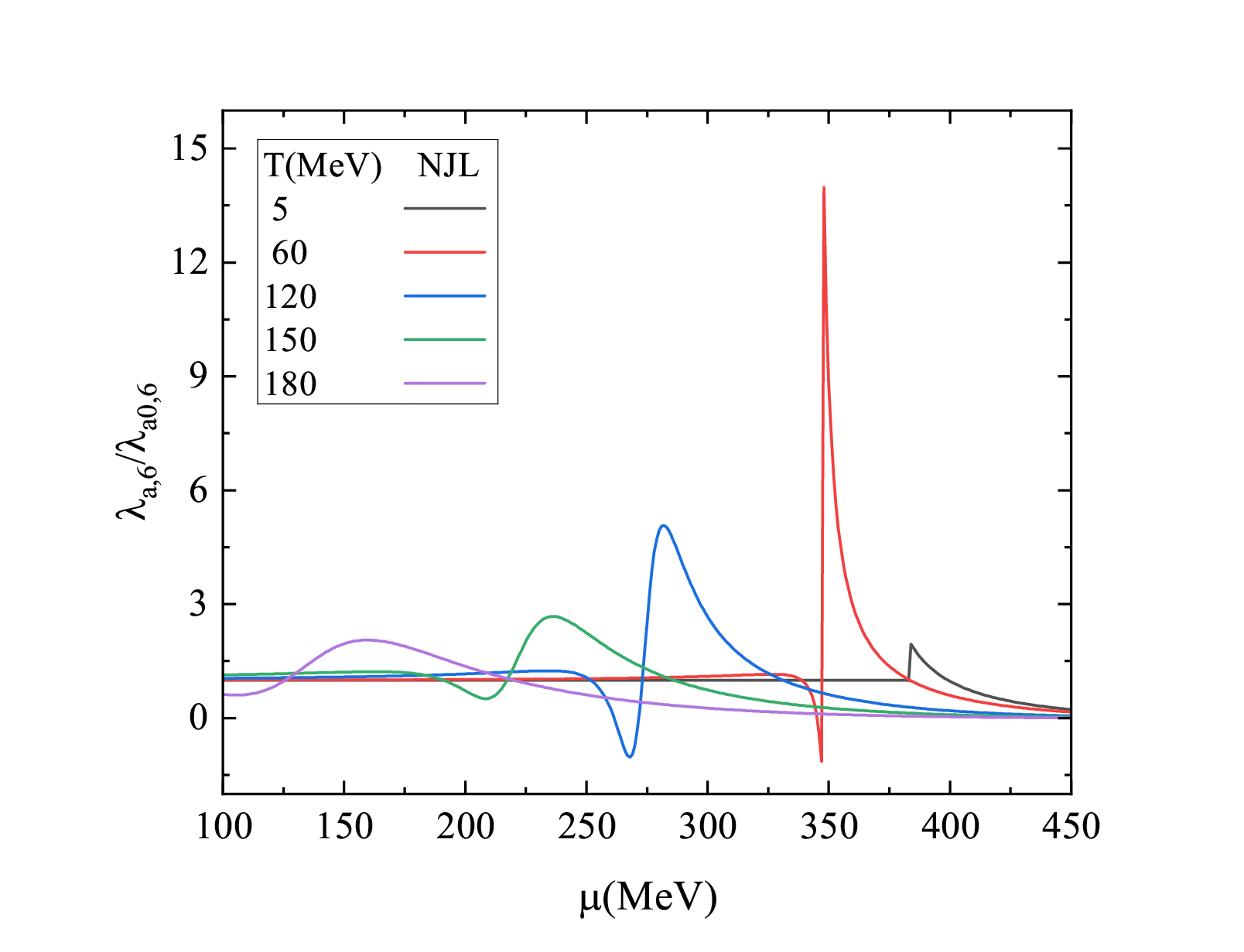}
\end{minipage}
\caption{The normalized axion sextic self-coupling constant $\lambda_{a,6}$ as the function of temperature at different quark chemical potentials (upper panel) 
and as the function of quark chemical potential at different temperatures (lower panel), obtained in the QM model. $\lambda_{a0,6}$ is the vacuum value of $\lambda_{a,6}$ 
calculated in the NJL model.}
\label{fig:sexticNJL}
\end{figure}

\begin{figure}[!htbp]
\centering
\begin{minipage}[h]{0.45\textwidth}
\centering
\includegraphics[width=1.0\textwidth]{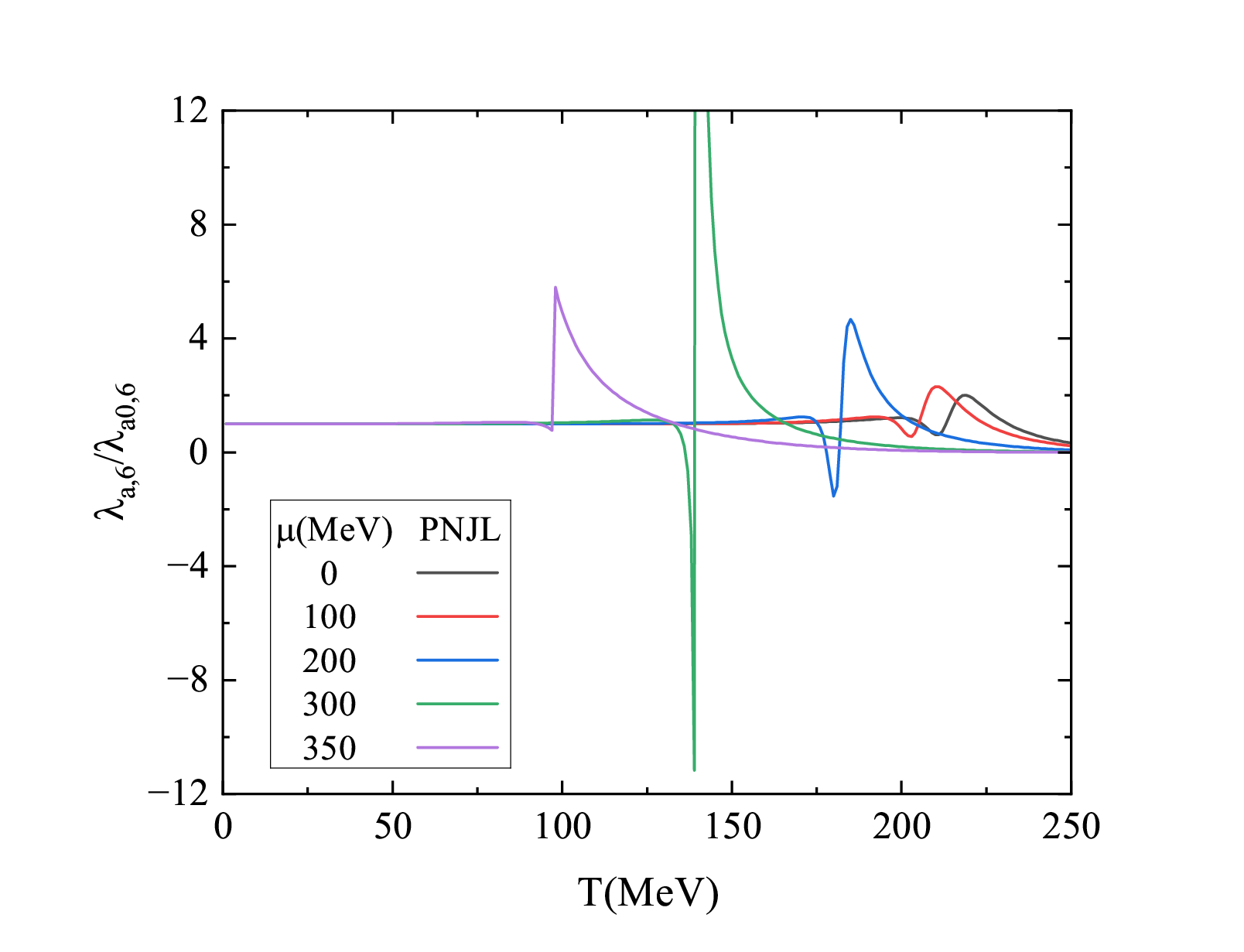}
\end{minipage}
\begin{minipage}[h]{0.45\textwidth}
\centering
\includegraphics[width=1.0\textwidth]{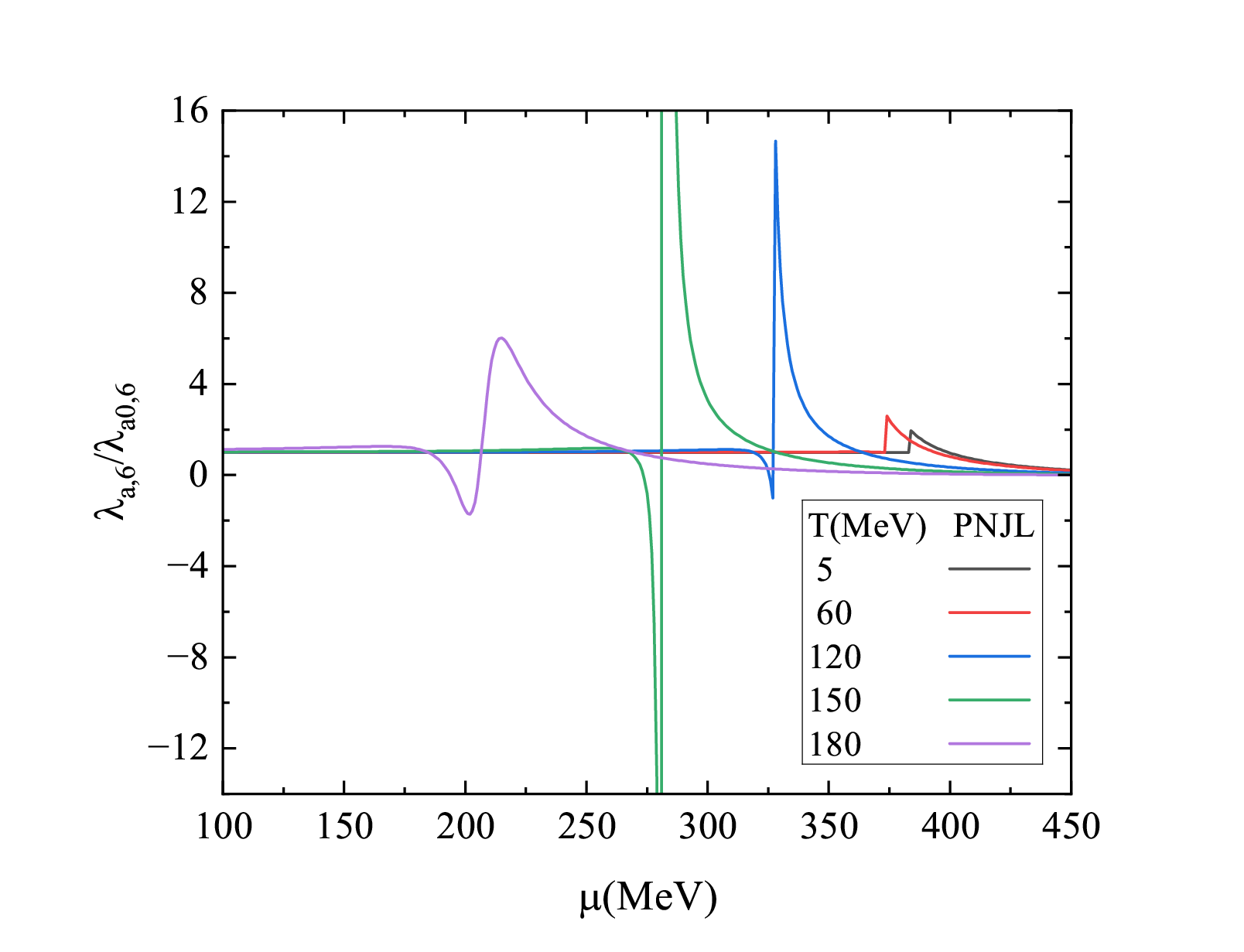}
\end{minipage}
\caption{Same as Fig. \ref{fig:b4NJL}, but obtained in the PNJL model.}
\label{fig:sexticPNJL}
\end{figure}

\section{The (P)NJL model with the $\theta$ term}

The Lagrangian density for the two-flavor NJL model at finite quark chemical potential $\mu$ is given by
\begin{equation}
\mathcal{L}_{NJL} = \bar{q} \left( i \gamma^\mu \partial_\mu + \mu \gamma_0 - m_0 \right) q + \mathcal{L}_{\text{int}}
\end{equation}
where $q = (u, d)^T$ represents the quark field doublet in flavor space, and $m_0$ is the degenerate current quark mass. 
The interaction Lagrangian $\mathcal{L}_{\text{int}}$ is defined as
\begin{equation}
\begin{split}
\mathcal{L}_{int}=G_1\left[ \left( \bar{q}\tau _aq \right) \left( \bar{q}\tau _aq \right) +\left( \bar{q}i\tau _a\gamma _5q \right) \left( \bar{q}i\tau _a\gamma _5q \right) \right] 
\\+8G_2\left[ e^{i\theta}\det \left( \bar{q}_Rq_L \right) +e^{-i\theta}\det \left( \bar{q}_Lq_R \right) \right], 
\end{split}
\end{equation}
where $\tau_a$ ($a = 1,2,3$) denote the Pauli matrices in flavor space. The first term is the usual four-quark interactions 
with the global $SU_L(2) \times SU_R(2)$ chiral symmetry. The second is the 't Hooft interaction, which explicitly breaks the global $U_A(1)$ 
symmetry. The $\theta$ angle is introduced via the $U_A(1)$ transformation.

Within the mean-field approximation, the full thermodynamic potential of the NJL model at one-loop order is given by 
\begin{equation}
\Omega_{\text{NJL}}=\Omega_{mf}+\Omega_{\text{q}}^{\text{vac}}+\Omega_{\text{q}}^{T}
\end{equation}
where
\begin{equation}
\begin{aligned}
\Omega_{mf} &= -G_2 \bigl( \eta^2 - \sigma^2 \bigr) \cos \theta + G_1 \bigl( \eta^2 + \sigma^2 \bigr) \\
&\quad - 2G_2 \sigma \eta \sin \theta, \label{eq:njlmf}
\end{aligned}
\end{equation}
and
\begin{align}
\Omega_{\text{q}}^{\text{vac}} &= -2N_c\sum_{f}\int_{0}^{\Lambda} {\frac{d\boldsymbol{p}}{(2\pi) ^3}}E_p^{f}, \label{eq:njlqv}
\end{align}
\begin{align}
\Omega_{\text{q}}^{T} &= -2N_cT\sum_{f}\int{\frac{d\boldsymbol{p}}{(2\pi) ^3} \left\{\ln \left[ 1 + e^{-( E_p^{f} -\mu) /T} \right] \right.} \notag \\
&\quad \left. + \ln [ 1 + e^{-( E_p^{f} +\mu) /T} ] \right\}, \label{eq:njlqT}
\end{align}
with $\sigma=\left<\bar{q}q\right>$ and $\eta =\left< \bar{q}i\gamma _5q \right>$. 
The dispersion relation takes the form 
\begin{equation}
E_p^f=\sqrt{p^2+M_f^2}, 
\end{equation}
where
\begin{equation}
\begin{split}
M_f^2&=\left( m_0+\alpha _0 \right) ^2+\beta _0^2 \\
\alpha_0 &=-2\left( G_1+G_2\cos \theta \right) \sigma +2G_2\eta \sin \theta \\
\beta _0 &=-2\left( G_1-G_2\cos \theta \right) \eta +2G_2\sigma \sin \theta.  
\end{split}
\end{equation}
We take the same model parameters as in \cite{Lu:2018ukl}, where $m_0=6$ MeV, $\varLambda=590$ MeV, and 
$G_0=2.435/\varLambda^2$, with $G_1=(1-c)G_0$, $G_2=cG_0$, and $c=0.2$. The parameters are fixed by matching  the empirical hadronic 
observables. The scalar condensate $\sigma$ and pseudoscalar condensate $\eta$ are determined by the gap equations
\begin{equation}
\frac{\partial \Omega_{\text{NJL}}}{\partial \sigma}=\frac{\partial \Omega_{\text{NJL}}}{\partial \eta}=0.
\end{equation}

The PNJL model can be constructed by incorporating the Polyakv loop dynamics to the NJL formalism. Here we only present the 
mean field potential of PNJL, which reads  
\begin{equation}
\Omega_{\text{PNJL}}=\Omega_{\text{mf}}+\Omega_{\text{PNJL}}^q+\mathcal{U}\left( \varPhi ,\bar{\varPhi} ,T \right)
\end{equation}
where
\begin{equation}
\Omega_{\text{PNJL}}^q=-2N_c\sum_{f}\int_{0}^{\Lambda}\frac{d\boldsymbol{p}}{(2\pi)^3}E_{p}^{f}-2T\sum_{f}\int\frac{d\boldsymbol{p}}{(2\pi)^3}(g_f^{+}+g_f^{-}). \label{19}
\end{equation}
We adopt the same Polyakov loop potential $\mathcal{U}\left( \varPhi ,\bar{\varPhi} ,T \right)$ \eqref{eq:plpt}.  
The condensates $\sigma$, $\eta$, as well as the Polyakov loop variable $\Phi$ and its conjugate $\bar{\Phi}$, are determined by 
the gap equations
\begin{equation}
\frac{\partial \Omega_{\text{PNJL}}}{\partial \sigma}=\frac{\partial \Omega_{\text{PNJL}}}{\partial \eta}=\frac{\partial \Omega_{\text{PNJL}}}{\partial \Phi} =\frac{\partial \Omega_{\text{PNJL}}}{\partial \bar{\Phi}}=0.
\end{equation}

\section{Cumulant $b_4$ and sextic self-coupling $\lambda_{a,6}$ in (P)NJL}

Figures \ref{fig:b4NJL} and \ref{fig:b4PNJL} show the $T$-$\mu$ dependence of the cumulant $b_4$ obtained within the NJL and PNJL 
formalisms, respectively. Figures \ref{fig:sexticNJL} and \ref{fig:sexticPNJL} are the corresponding sextic self-couplings calculated 
under the same conditions. We see that critical behaviors of $b_4$ and $\lambda_{a,6}$ are qualitatively consistent with that computed  
using the (P)QM model and their asymptotic values ar high temperature and/or density are quantitatively consistent with the QM results .      

\begin{figure}[!htbp]
\centering
\begin{minipage}[h]{0.45\textwidth}
\centering
\includegraphics[width=1.0\textwidth]{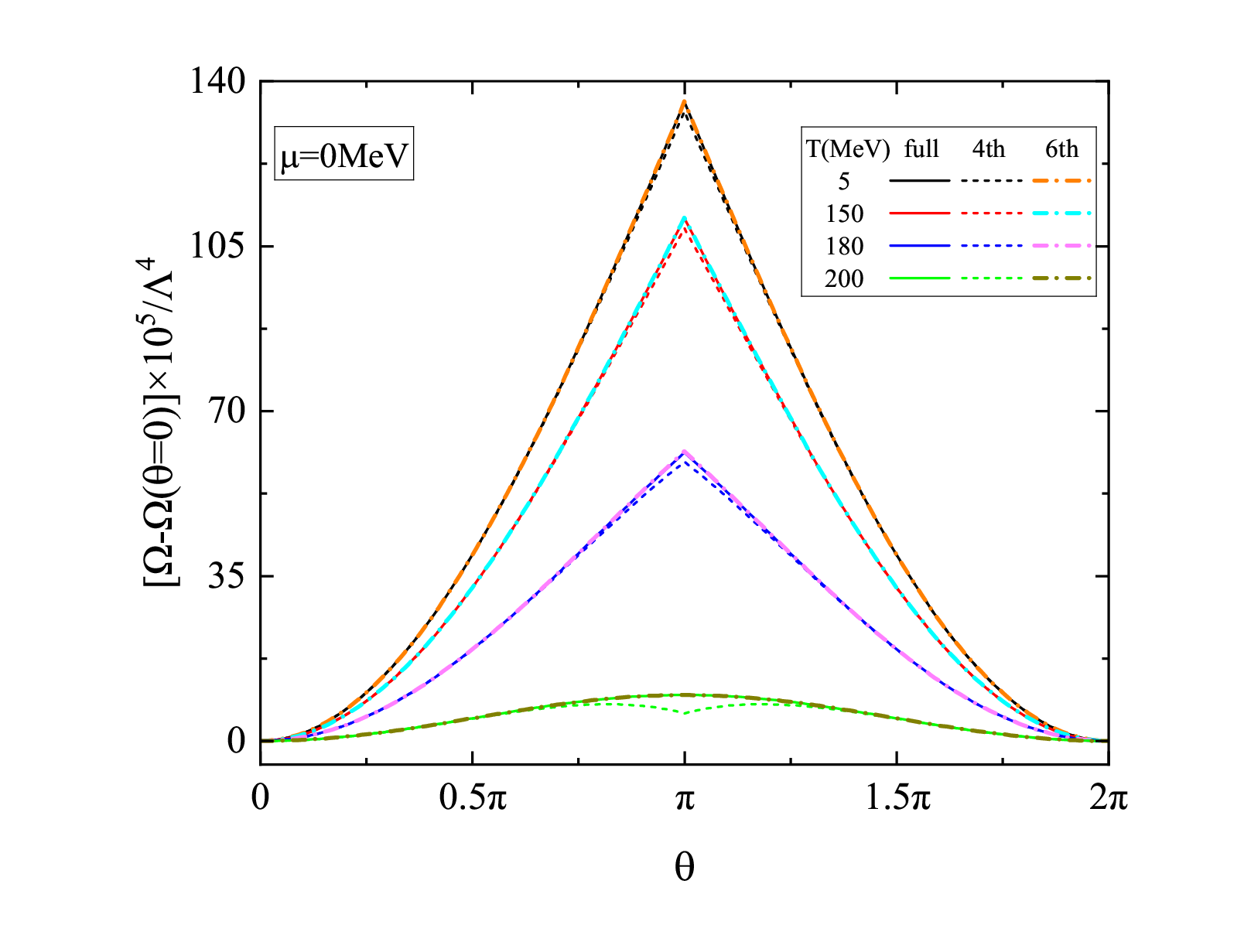}
\end{minipage}
\begin{minipage}[h]{0.45\textwidth}
\centering
\includegraphics[width=1.\textwidth]{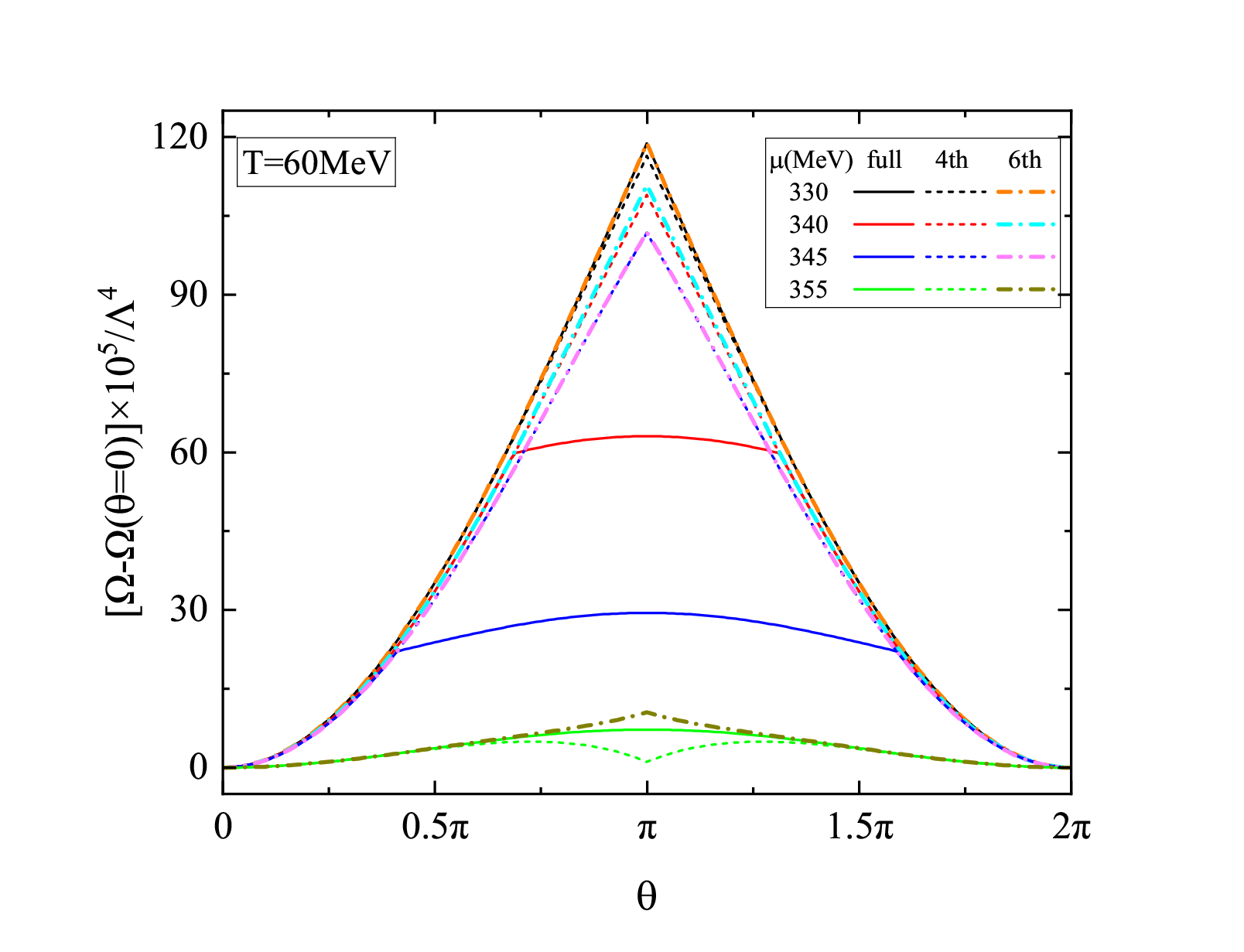}
\end{minipage}
\caption{Full potential $\Omega(\theta)$ and its fourth- and sixth-order Taylor expansions within the NJL formalism: 
upper panel for $\mu=0$ at different temperatures and lower panel for $T=60~\text{MeV}$ at different chemical potentials.}
\label{fig:taylorNJL}
\end{figure}

\begin{figure}[!htbp]
\centering
\begin{minipage}[h]{0.45\textwidth}
\centering
\includegraphics[width=1.0\textwidth]{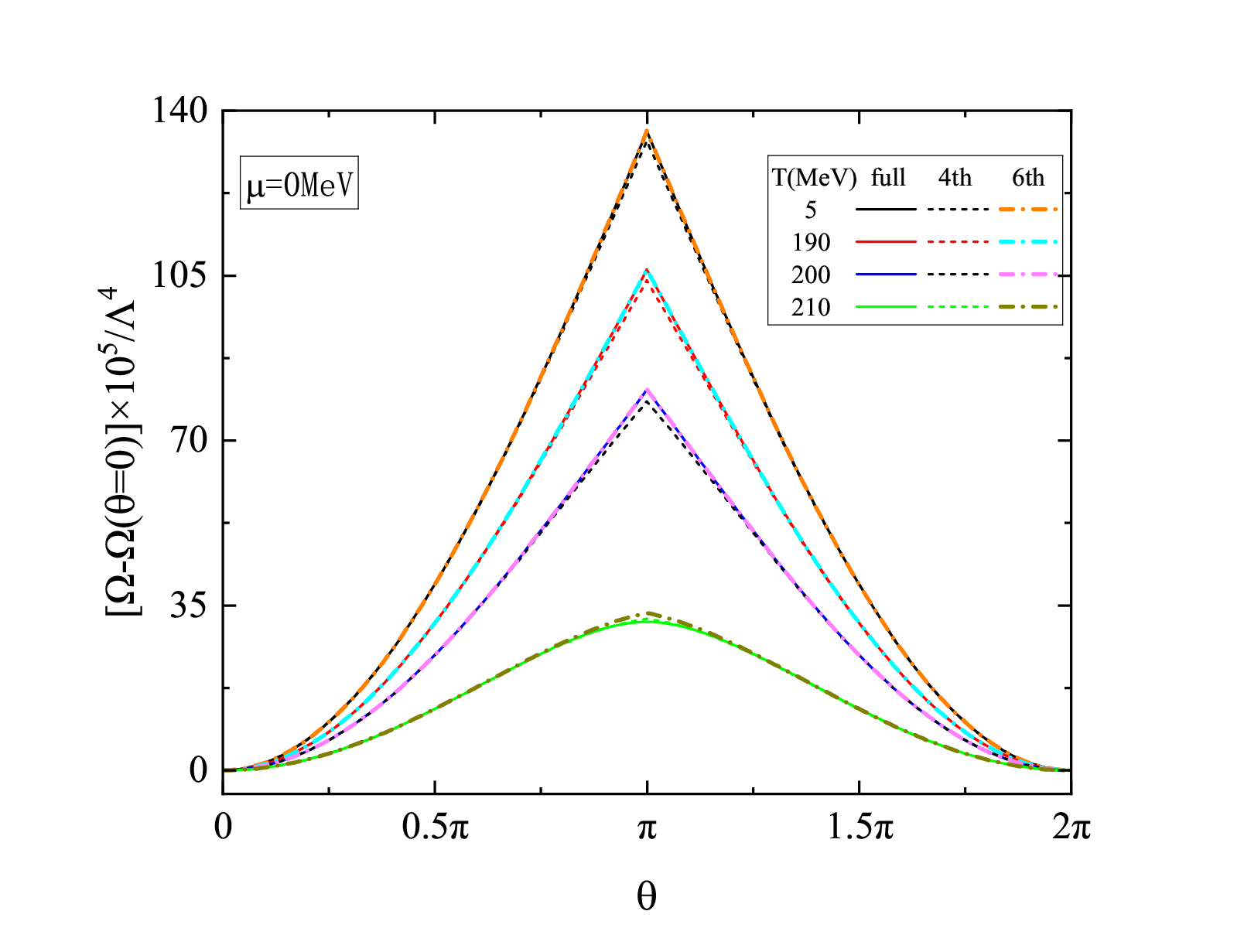}
\end{minipage}
\begin{minipage}[h]{0.45\textwidth}
\centering
\includegraphics[width=1.0\textwidth]{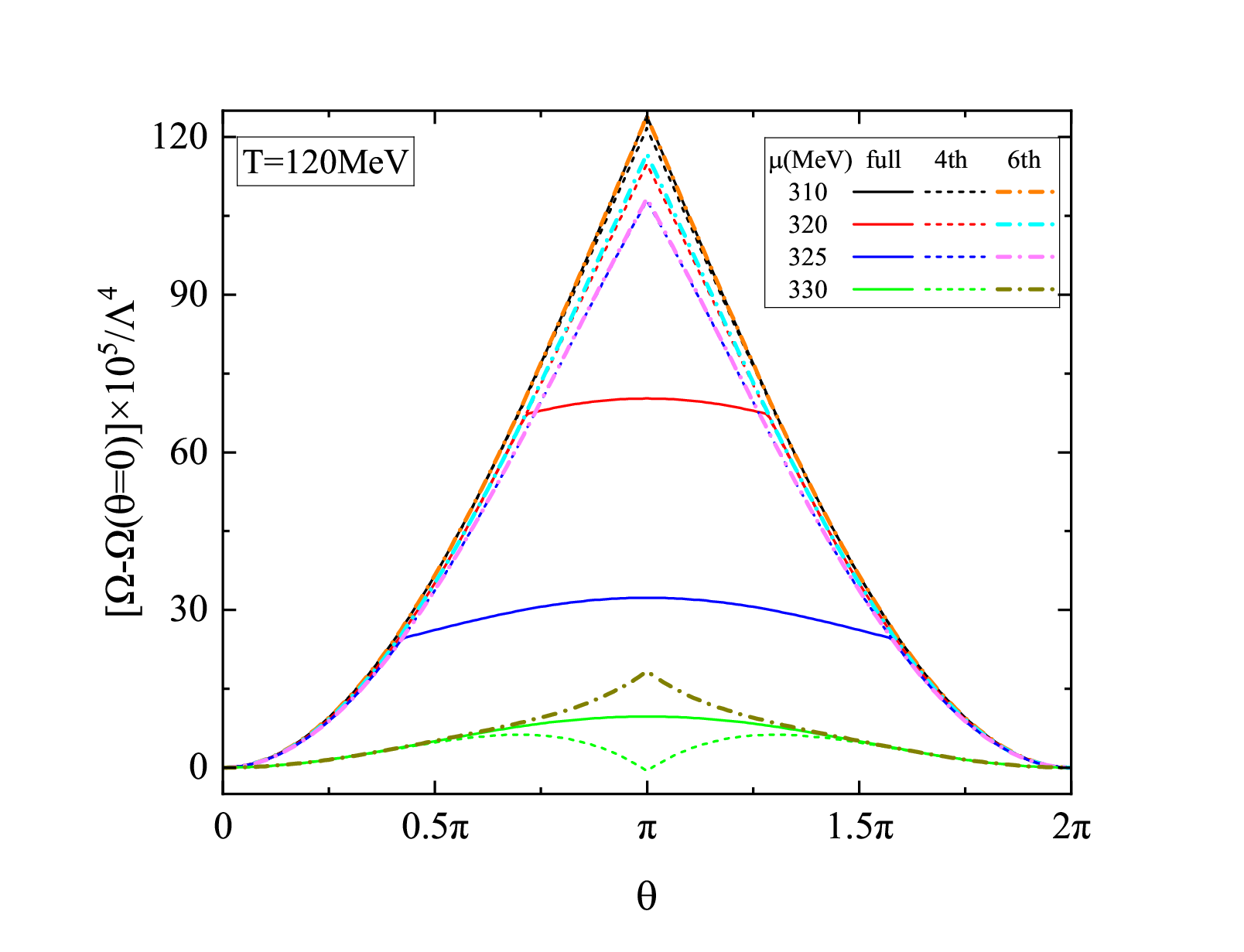}
\end{minipage}
\caption{Full potential $\Omega(\theta)$ and its fourth- and sixth-order Taylor expansions within the PNJL formalism: 
upper panel for $\mu=0$ at different temperatures and lower panel for $T=120~\text{MeV}$ at different chemical potentials.}
\label{fig:taylorPNJL}
\end{figure}

\section{Range of accuracy of the Taylor expansion approximation of the potential with respect to $\theta$ in (P)NJL}
 
The full thermal potential at the mean-field level and the corresponding fourth- and sixth-order Taylor expansions as functions 
of $\theta$ at different temperatures and chemical potentials obtained within the NJL and PNJL models are shown in Figs. \ref{fig:taylorNJL} 
and \ref{fig:taylorPNJL}, respectively. The range of accuracy of the Taylor expansion approximation under various conditions 
is similar to that calculated in the (P)QP model.

\end{document}